\documentclass[11 pt]{article}
\usepackage[mathscr]{eucal}
\usepackage{amssymb,latexsym}
\usepackage{verbatim}
\usepackage{amsmath}
\usepackage{amsthm}
\usepackage{enumerate}
\usepackage{authblk}
\usepackage{color}
\usepackage{multicol}
\usepackage{tikz}
\usepackage{xypic}
\usepackage{caption}

\usetikzlibrary{snakes}

\newtheorem{thm}{Theorem}[section]
\newtheorem{lem}[thm]{Lemma}
\newtheorem{Def}[thm]{Definition}
\newtheorem{prop}[thm]{Proposition}

\renewcommand\l{\lambda}

\renewcommand\S{\Sigma}

\newcommand\s{\sigma}

\newcommand\e{\varepsilon}
\renewcommand\b{\beta}

\renewcommand\l{\lambda}
\newcommand\g{\gamma}

\renewcommand\a{\alpha}

\newcommand\beq{\begin{equation}}
\newcommand\eeq{\end{equation}}
\newcommand\ben{\begin{enumerate}}
\newcommand\een{\end{enumerate}}
\newcommand\bit{\begin{itemize}}
\newcommand\eit{\end{itemize}}

\newcommand{\R}{\mathbb R}

\newcommand{\ov}{\overline}
\newcommand{\ms}{\mathscr}

\newcommand{\ext}{\text{{\rm ext}}}
\newcommand{\safe}{\text{{\rm safe}}}
\newcommand{\unsafe}{\text{{\rm unsafe}}}

\newcommand{\pd}{\partial}

\newcommand{\mc}{\mathcal}

\newcommand{\C}{\mathbb{C}}

\def\undertilde#1{\mathord{\vtop{\ialign{##\crcr
   $\hfil\displaystyle{#1}\hfil$\crcr\noalign{\kern1.5pt\nointerlineskip}
   $\hfil\tilde{}\hfil$\crcr\noalign{\kern1.5pt}}}}}

\newcounter{mnotecount}

\title{The Big Bang is a Coordinate Singularity for $k = -1$ Inflationary FLRW Spacetimes
}

\author{Eric Ling\footnote{eling@math.miami.edu}}
\affil{Department of Mathematics
\\ KTH Royal Institute of Technology }

\begin{document}
\date{}
\maketitle
\vspace{.2in}

\begin{abstract} 
We show that the big bang is  a coordinate singularity for a large class of $k = -1$ inflationary FLRW spacetimes which we have dubbed `Milne-like.' By introducing a new set of coordinates, the big bang appears as a past boundary of the universe where the metric is no longer degenerate -- a result which has already been investigated in the context of vacuum decay \cite{ColemanFrank}. We generalize their results and approach the problem from a more mathematical perspective. Similar to how investigating the geometrical properties of the $r = 2m$ event horizon in Schwarzschild led to a better understanding of black holes, we believe that investigating the geometrical properties of the big bang coordinate singularity for Milne-like spacetimes could lead to a better understanding of cosmology. We  show how the mathematics of these spacetimes may help illuminate certain issues associated with dark energy, dark matter, and the universe's missing antimatter.
\end{abstract}

\newpage

\tableofcontents

\newpage

\section{Introduction}

In this paper we show that the big bang is a coordinate singularity for a large class of $k = -1$ inflationary FLRW spacetimes which we have dubbed `Milne-like.' This may seem surprising and one may justifiably ask: don't the singularity theorems imply that this cannot happen? In appendix \ref{singularity theorem appendix} we show why the singularity theorems don't apply to inflationary spacetimes. Indeed Milne-like spacetimes can almost always be used as counterexamples.

The coordinate singularity appearing in Milne-like spacetimes may offer a new geometrical perspective of our universe. We believe that understanding the geometry of the coordinate singularity for Milne-like spacetimes could lead to a better understanding of cosmology. One of the goals of this paper is to demonstrate how the mathematics of these spacetimes may help illuminate certain issues associated with dark energy, dark matter, and the universe's missing antimatter. Our solution to the antimatter problem is similar to the solution presented in \cite{TBF}.

We remark a previous time when understanding the geometry of a coordinate singularity led to advances in theoretical physics. The $r = 2m$ event horizon in Schwarzschild is a coordinate singularity. Understanding the geometry of event horizons led to Hawking's area theorem for black holes \cite{HE, Wald, CDGH} which played a pivotal role in the development of black hole thermodynamics \cite{Wald}.

Throughout this paper we set constants $G = c = \hbar = 1$. Our signature convention is $(-,+,+,+)$.

\medskip
\medskip

\subsection*{Acknowledgments}
The author thanks Greg Galloway, Paul Tod, Bob Wald, Hugh Bray, Gary Horowitz, \linebreak Gerald Folland, Orlando Alvarez, Michel Mizony, Bob Klauber, and Torsten Asselmeyer-Maluga for helpful comments and discussions.

\medskip
\medskip

\subsection{Summary of results}

\medskip

In this section we give a brief summary of our results. We highlight four main points.

\begin{itemize}

\item[(1)] The big bang is a coordinate singularity for Milne-like spacetimes.

\item[(2)] The geometric solution to the horizon problem.

\item[(3)] The cosmological constant appears as an initial condition.

\item[(4)] Lorentz invariance and its implications for dark matter and antimatter.

\end{itemize}

\medskip
\medskip

\noindent{\bf (1) The big bang is a coordinate singularity for Milne-like spacetimes.}

\medskip
\medskip

The Milne universe is the $k = -1$ FLRW spacetime with manifold $M = (0,\infty) \times \R^3$ and metric
\begin{equation}
g \,=\, -d\tau^2 + a^2(\tau)\big[dR^2 + \sinh^2(R)(d\theta^2 + \sin^2\theta d\phi^2 ) \big]
\end{equation}
and scale factor $a(\tau) = \tau$.
Here we are using hyperbolic coordinates $(\tau, R, \theta, \phi)$. By introducing coordinates $(t,r,\theta,\phi)$ via $t = \tau \cosh(R)$ and $r = \tau \sinh(R)$, the metric is 
\begin{equation}
g \,=\, -dt^2 + dr^2 + r^2(d\theta^2 + \sin^2\theta d\phi^2)
\end{equation}
which is just the Minkowski metric. Notice that $\tau = 0$ corresponds to the lightcone $t = r$ at the origin $\ms{O}$. Hence the big bang is a coordinate singularity for the Milne universe.

A Milne-like spacetime is a $k = -1$ FLRW spacetime with scale factor assumed to satisfy $a(\tau) = \tau + o(\tau^{1 + \e})$ for small $\tau$. Letting $\e = 1/2$ shows that this condition is satisfied for any Taylor expansion $a(\tau) = \tau + c_2\tau^2 + c_3\tau^3 + \dotsb$. Since this is a limiting condition, Milne-like spacetimes can include an inflationary era, a radiation-dominated era, a matter-dominated era, and a dark energy-dominated era. Therefore they can model the dynamics of our universe. We introduce coordinates $(t,r,\theta, \phi)$ via $t = b(\tau) \cosh(R)$ and $r = b(\tau)\sinh(R)$ where $b$ satisfies $b' = b/a$. Putting $\Omega = 1/b' = a/b$, the metric is
\begin{equation}\label{conformal metric intro eq}
g \,=\, \Omega^2(\tau)\big[-dt^2 + dr^2 + r^2(d\theta^2 + \sin^2\theta d\phi^2) \big].
\end{equation} 
The condition $a(\tau) = \tau + o(\tau^{1 + \e})$ implies $0 < \Omega(0) < +\infty$ where $\Omega(0) :=\lim_{\tau \to 0} \Omega(\tau)$.  Therefore there is no degeneracy at $\tau = 0$ in these coordinates. Thus the big bang is a coordinate singularity for Milne-like spacetimes. See section \ref{milne-like spacetimes section} for the full proof.

\medskip
\medskip

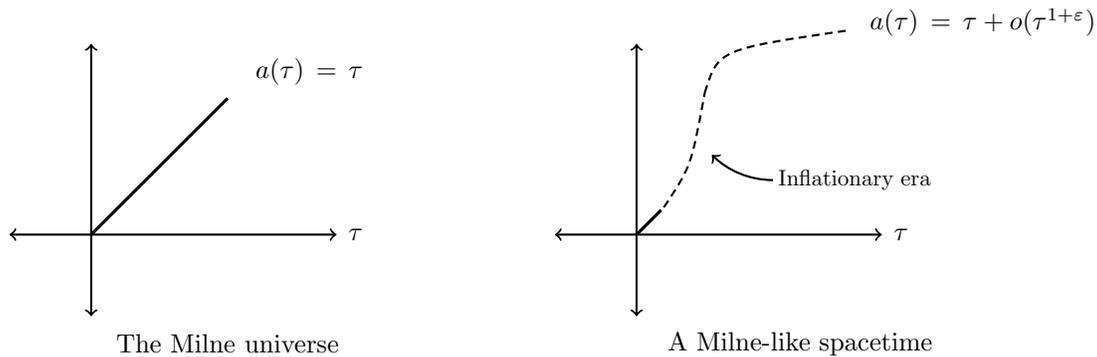
\begin{figure}[h]
\[
\begin{tikzpicture}[scale = .725]

\draw [<->,thick] (-12,-2.5) -- (-12,2.5);
\draw [<->,thick] (-13.5,-1) -- (-7.5,-1);
\draw [very thick] (-12,-1) -- (-9.5,1.5);

\draw (-7.15, -1) node {\small{$\tau$}};

\draw  (-8, 2) node {\small{$a(\tau) \,=\, \tau$}};

\draw (-9.5,-3) node {\small{The Milne universe}};


\draw [<->,thick] (-2,-2.5) -- (-2,2.5);
\draw [<->,thick] (-3.5,-1) -- (2.5,-1);

\draw (2.85, -1) node {\small{$\tau$}};

\draw [very thick] (-2,-1) -- (-1.55,-0.55);
\draw [densely dashed, thick] (-1.5, -0.5) .. controls (-1,.25).. (-.75,1.6);
\draw [densely dashed, thick] (-.75, 1.6) .. controls (-.5,2.4).. (1.9,2.75);

\draw  (4.35, 2.9) node {\small{$a(\tau) \,=\, \tau + o(\tau^{1 +\e})$}};

\draw [->] [thick] (0.5,0) arc [start angle=-90, end angle=-135, radius=45pt];
\draw (2.0,0) node [scale = .85]{\small{Inflationary era}};

\draw (1.0,-3) node {\small{A Milne-like spacetime}};

\end{tikzpicture}
\]
\captionsetup{format=hang}
\caption{\small{Left: The scale factor for the Milne universe. Right: The scale factor for a Milne-like spacetime modeling the dynamics of inflation.}}\label{milne universe and milne-like scale factor figure}
\end{figure}

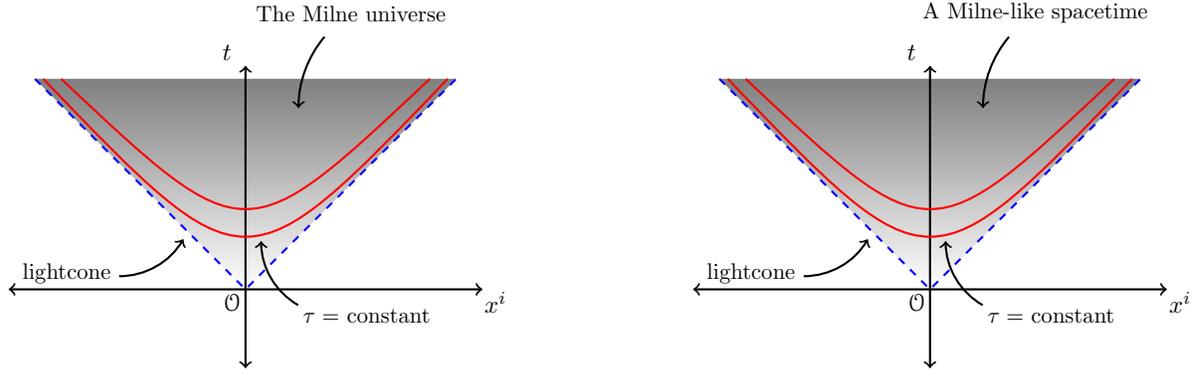
\begin{figure}[h]
\[
\begin{tikzpicture}[scale = .7]

\shadedraw [dashed, thick, blue](-4,2) -- (0,-2) -- (4,2);

\draw [<->,thick] (0,-3.5) -- (0,2.25);
\draw [<->,thick] (-4.5,-2) -- (4.5,-2);

\draw (-.35,2.5) node [scale = .85] {$t$};
\draw (4.75, -2.25) node [scale = .85] {$x^i$};
\draw (-.25,-2.25) node [scale = .85] {$\ms{O}$};


\draw [->] [thick] (1.5,2.8) arc [start angle=140, end angle=180, radius=60pt];
\draw (2.0,3.25) node [scale = .85]{\small{The Milne universe}};

\draw [->] [thick] (-2.4,-1.75) arc [start angle=-90, end angle=-30, radius=40pt];
\draw (-3.4,-1.7) node [scale = .85] {\small lightcone};


\draw [thick, red] (-3.84,2) .. controls (0,-2) .. (3.84,2);
\draw [thick, red] (-3.5,2) .. controls (0, -1.3).. (3.5,2);

\draw [->] [thick]  (1,-2.3) arc [start angle=-120, end angle=-180, radius=40pt];
\draw (2.3,-2.5) node [scale = .85] {\small{$\tau =$ constant }};


\shadedraw [dashed, thick, blue](9,2) -- (13,-2) -- (17,2);

\draw [<->,thick] (13,-3.5) -- (13,2.25);
\draw [<->,thick] (8.5,-2) -- (17.5,-2);

\draw (12.65,2.5) node [scale = .85] {$t$};
\draw (17.75, -2.25) node [scale = .85] {$x^i$};
\draw (12.75,-2.25) node [scale = .85] {$\ms{O}$};


\draw [->] [thick] (14.5,2.8) arc [start angle=140, end angle=180, radius=60pt];
\draw (15.0,3.25) node [scale = .85]{\small{A Milne-like spacetime}};

\draw [->] [thick] (10.6,-1.75) arc [start angle=-90, end angle=-30, radius=40pt];
\draw (9.6,-1.7) node [scale = .85] {\small lightcone};


\draw [thick, red] (9.16,2) .. controls (13,-2) .. (16.84,2);
\draw [thick, red] (9.5,2) .. controls (13, -1.3).. (16.5,2);

\draw [->] [thick]  (14,-2.3) arc [start angle=-120, end angle=-180, radius=40pt];
\draw (15.3,-2.5) node [scale = .85] {\small{$\tau =$ constant }};

\end{tikzpicture}
\]
\captionsetup{format=hang}
\caption{\small{Left: the Milne universe sits inside the future lightcone at the origin $\ms{O}$ of Minkowsi space. Right: a Milne-like spacetime sits inside the future lightcone at the origin $\ms{O}$ of a spacetime conformal to Minkowski space. In both caes the spacetime extends through the lightcone at the origin $\ms{O}$.}}\label{milne universe and milne-like figure}
\end{figure}

\medskip
\medskip

Figure \ref{milne universe and milne-like figure} shows how the big bang is a coordinate singularity for the Milne universe and Milne-like spacetimes. In both cases the lightcone at the origin $\ms{O}$ acts as a past boundary of the universe. It separates our universe from the extension.

The discussion so far has been informal. In section \ref{definition section} we give precise mathematical definitions to what we mean by coordinate and curvature singularities. We then apply these definitions in section \ref{coord singularity section} to show the big bang is a coordinate singularity for Milne-like spacetimes. In section \ref{C2 section} we show  Milne-like spacetimes admit no curvature singularities provided the second derivative of the scale factor satisfies an asymptotic condition.

\medskip
\medskip

\newpage

\noindent{\bf (2) The geometric solution to the horizon problem.}

\medskip
\medskip

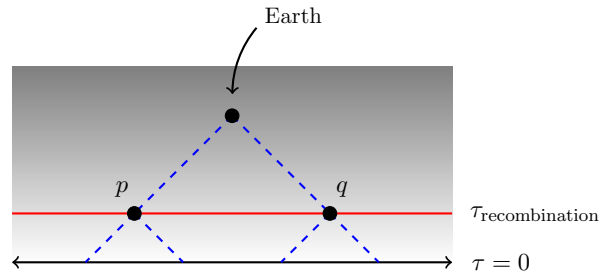
\begin{figure}[h]
\[
\begin{tikzpicture}[scale = 0.65]
\shade(-4.5,2) -- (-4.5,-2) -- (4.5,-2) -- (4.5,2);

\draw [<->,thick] (-4.5,-2) -- (4.5,-2);

\draw [->] [thick] (.5,2.8) arc [start angle=140, end angle=180, radius=60pt];
\draw (1.25,3.05) node [scale = .85]{\small{Earth}};

\draw (5.5, -2) node [scale = .85] {$\tau = 0$};

\draw [thick, red] (-4.5,-1) -- (4.5,-1);

\draw [dashed, thick, blue] (0,1) -- (2,-1);
\draw [dashed, thick, blue] (0,1) -- (-2,-1);

\draw [dashed, thick, blue] (-2,-1) -- (-1,-2);
\draw [dashed, thick, blue] (-2,-1) -- (-3,-2);

\draw [dashed, thick, blue] (2,-1) -- (3,-2);
\draw [dashed, thick, blue] (2,-1) -- (1,-2);

\draw (6.2,-1) node [scale = .85] {$\tau_\text{recombination}$  };

\node [scale = .50] [circle, draw, fill = black] at (-2,-1)  {};
\node [scale = .50] [circle, draw, fill = black] at (2,-1)  {};
\node [scale = .50] [circle, draw, fill = black] at (0,1)  {};
\draw (-2.25,-.5) node [scale =.85] {$p$};
\draw (2.25,-.5) node [scale =.85] {$q$};

\end{tikzpicture}
\]
\captionsetup{format=hang}
\caption{\small{The horizon problem. Without inflation the past lightcones of $p$ and $q$ don't intersect.}}
\label{horizon problem intro fig}
\end{figure}

We briefly recall the horizon problem in cosmology. It is the main motivating reason for inflationary theory \cite[page 208]{WeinbergCos}. The problem comes from the uniform temperature of the CMB radiation. From any direction in the sky, we observe the CMB temperature as 2.7 K. The uniformity of this temperature is puzzling: if we assume the universe exists in a radiation-dominated era all the way down to the big bang (i.e. no inflation), then the points $p$ and $q$ on the surface of last scattering don't have intersecting past lightcones. So  how can the CMB temperature be so uniform if $p$ and $q$ were never in causal contact in the past? 
See Figure \ref{horizon problem intro fig}.

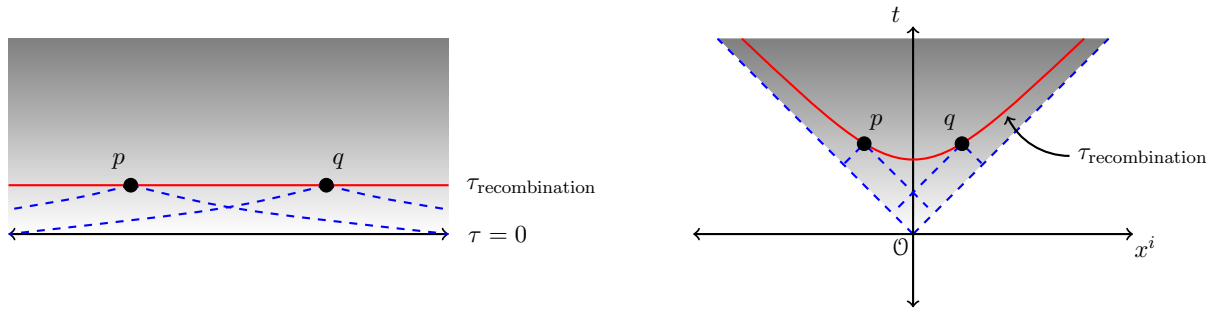
\begin{figure}[h]
\[
 \begin{tikzpicture}[scale = 0.65]

\shade(-8.5,2) -- (-8.5,-2) -- (0.5,-2) -- (0.5,2);

\draw [<->,thick] (-8.5,-2) -- (0.5,-2);

\draw (1.5, -2) node [scale = .85] {$\tau = 0$};

\draw [thick, red] (-8.5,-1) -- (0.5,-1);

\draw [dashed, thick, blue] (-6,-1) .. controls (-4,-1.5) .. (0.5,-2);
\draw [dashed, thick, blue] (-6,-1) .. controls (-7,-1.25) .. (-8.5,-1.5);
\draw [dashed, thick, blue] (-2,-1) .. controls (-1,-1.25) .. (0.5,-1.5);
\draw [dashed, thick, blue] (-2,-1) .. controls (-4,-1.5) .. (-8.5,-2);

\node [scale = .50] [circle, draw, fill = black] at (-6,-1)  {};
\node [scale = .50] [circle, draw, fill = black] at (-2,-1)  {};

\draw (-6.25,-.5) node [scale =.85] {$p$};
\draw (-1.75,-.5) node [scale =.85] {$q$};

\draw (2.2,-1) node [scale = .85] {$\tau_\text{recombination}$  };



\shadedraw [white] (6,2) -- (10,-2) -- (14,2);
\draw [dashed, thick, blue] (10,-2) -- (14,2);
\draw [dashed, thick, blue] (10,-2) -- (6,2);

\draw [<->,thick] (10,-3.5) -- (10,2.25);
\draw [<->,thick] (5.5,-2) -- (14.5,-2);

\draw (9.65,2.5) node [scale = .85] {$t$};
\draw (14.75, -2.25) node [scale = .85] {$x^i$};

\draw (9.75,-2.25) node [scale = .85] {$\ms{O}$};

\draw [thick, red] (6.5,2) .. controls (10, -1.3).. (13.5,2);

\draw [dashed, thick, blue](8.6,-.55) -- (9,-0.15) -- (10.3,-1.45);
\draw [dashed, thick, blue](9.7,-1.45) -- (11,-0.15) -- (11.4,-.55);

\node [scale = .50] [circle, draw, fill = black] at (9,-.15)  {};
\node [scale = .50] [circle, draw, fill = black] at (11,-.15)  {};

\draw (9.25,0.3) node [scale =.85] {$p$};
\draw (10.75,0.3) node [scale =.85] {$q$};

\draw [->] [thick]  (13.2,-0.4) arc [start angle=-90, end angle=-155, radius=40pt];
\draw (14.7,-0.4) node [scale = .85] {$\tau_\text{recombination}$  };
\end{tikzpicture}
\]
\captionsetup{format=hang}
\caption{\small{An inflationary era widens the past lightcones to solve the horizon problem. This is depicted in the figure on the left. The figure on the right depicts the solution for Milne-like spacetimes in the conformal Minkowski coordinates.}}
\label{horizon problem solution intro fig}
\end{figure}

An inflationary era before the radiation-dominated era would allow for causal contact in the past. This is depicted in Figure \ref{horizon problem solution intro fig}. For a Milne-like spacetime the solution becomes apparent in the conformal Minkowski coordinates. Since the spacetime is conformal to Minkowski space, the lightcones are given by 45 degree angles. Therefore any two points $p$ and $q$ in a Milne-like spacetime will have past lightcones which intersect at some point above the origin $\ms{O}$.

\medskip
\medskip

\noindent{\bf (3) The cosmological constant appears as an initial condition.}

\medskip
\medskip

Here we show how the cosmological constant $\Lambda$ appears as an initial condition for Milne-like spacetimes. This may help explain the origin of $\Lambda$. If dark energy is really modeled by a cosmological constant and not by some other model (e.g. quintessence), then $\Lambda$ would have been fixed at the big bang.

The Einstein equations with a cosmological constant are
\begin{equation}
 R_{\mu\nu} - \frac{1}{2}R g_{\mu\nu} + \Lambda g_{\mu\nu} \,=\, 8\pi T_{\mu\nu}.
\end{equation}
Let $u = \pd / \pd \tau$ denote the four-velocity of the comoving observers and let $e$ be any unit spacelike orthogonal vector (its choice does not matter by isotropy). We define the energy density $\rho$ and pressure function $p$ in terms of the Einstein tensor: $\rho = \frac{1}{8\pi}G_{\mu\nu}u^\mu u^\nu$ and $p = \frac{1}{8\pi}G_{\mu\nu}e^\mu e^\nu$ where $G_{\mu\nu} = R_{\mu\nu} - \frac{1}{2}R g_{\mu\nu}$. 
 We define the \emph{normal} energy density $\rho_{\rm normal}(\tau)$ and \emph{normal} pressure function $p_{\rm normal}(\tau)$ in terms of the energy-momentum tensor $\rho_{\rm normal} = T_{\mu\nu}u^\mu u^\nu$ and $p_{\rm normal} = T_{\mu\nu}e^\mu e^\nu$.
Therefore $\rho = \rho_{\rm normal} + \Lambda/ 8\pi$ and $p = p_{\rm normal} - \Lambda/ 8\pi$.

If $\rho_{\rm normal} = p_{\rm normal} = 0$ (e.g. de Sitter), then the equation of state for the cosmological constant is fixed for all $\tau$.
\begin{equation}
\rho \, = \, - p \, =\,  \frac{\Lambda}{8\pi}.
\end{equation}

In section \ref{cosmo constant section} we show that this equation of state appears as an initial condition for Milne-like spacetimes. Specifically, we show if the scale factor satisfies $a''(\tau) = \a \tau + o(\tau)$, then 
\begin{equation}
\rho(0) \, = \, -p(0) \, = \, \frac{3}{8\pi} \a .
\end{equation}

Consequently, if $\rho_{\rm normal}(\tau)$ and $p_{\rm normal}(\tau) \to 0$ as $\tau \to 0$ (i.e. if the cosmological constant was the dominant energy source during the Planck era), then we have $\Lambda \,=\, 3\a.$ Note that $\a = a'''(0)$. Therefore we have
\begin{equation}
\Lambda \,=\, 3a'''(0).
\end{equation}
This provides a connection between the cosmological constant $\Lambda$ and the initial condition of the scale factor.

\medskip
\medskip

\noindent{\bf (4) Lorentz invariance and its implications for dark matter and antimatter.}

\medskip
\medskip

In section \ref{Lorentz invariance section} we show the Lorentz group $\text{L} = \text{O}(1,3)$ acts by isometries on Milne-like spacetimes at the origin $\ms{O}$. Therefore these spacetimes have a notion of Lorentz invariance. This follows because the Minkowski metric is Lorentz invariant and the conformal factor $\Omega$ appearing in equation (\ref{conformal metric intro eq}) is Lorentz invariant. Since Lorentz invariance plays a pivotal role in QFT (e.g. the field operators are constructed out of finite dimensional irreducible represenations of the Lorentz group \cite{Weinberg, Tung}), Milne-like spacetimes are a good background model if one wants to develop a quantum theory of cosmology with Lorentz invariance.

\medskip
\medskip

\noindent{\bf A possible dark matter particle?}

\medskip

There are two different kinds of symmetries in quantum theory. The local (gauge) symmetry group $\text{SU}(3) \times \text {SU}(2) \times \text{U}(1)$ of the standard model and the global spacetime symmetry group. 

The irreducible unitary representations of $\text{SL}(2,\C)$ (which is the simply connected double cover of the connected component of $\text{O}(1,3)$ containing the identity) are characterized by two different types of particles. See Theorem 10.9 in \cite{Tung}. The first class of particles is the \emph{principal series}. These particles are characterized by the parameter $\nu = -iw$ where $w$ is real and spin $j = 0, 1/2, 1, \dotsc$ The second class of particles is the \emph{complementary series}. These particles are characterized by the parameter $-1 \leq \nu \leq 1$ and spin $j = 0$. The terminology `principal' and `complementary' comes from the classification of irreducible unitary representations of semi-simple Lie groups \cite{Knapp}.

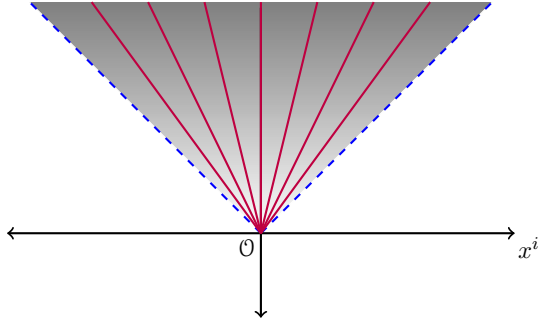
\begin{figure}[h]
\[
\begin{tikzpicture}[scale = 0.75]

\shadedraw [white] (-4.1,2.1) -- (0,-2) -- (4.1,2.1);
\draw [dashed, thick, blue] (0,-2) -- (4.1,2.1);
\draw [dashed, thick, blue] (0,-2) -- (-4.1,2.1);

\draw [<-,thick] (0,-3.5) -- (0,2.0);
\draw [<->,thick] (-4.5,-2) -- (4.5,-2);

\draw (4.75, -2.25) node [scale = .85] {$x^i$};
\draw (-.25,-2.25) node [scale = .85] {$\ms{O}$};
%

	\draw [thick, purple] (0,-2) -- (2,2.1);
	\draw [thick, purple] (0,-2) -- (3,2.1);
	\draw [thick, purple] (0,-2) -- (1,2.1);
	\draw [thick, purple] (0,-2) -- (-1,2.1);
	\draw [thick, purple] (0,-2) -- (-2,2.1);
	\draw [thick, purple] (0,-2) -- (-3,2.1);
	\draw [thick, purple] (0,-2) -- (0,2.1);

\end{tikzpicture}
\]
\captionsetup{format=hang}
\caption{\small{The comoving observers in a Milne-like spacetime. They all emanate from the origin $\ms{O}$.}}\label{comoving figure in intro}
\end{figure}

Since the comoving observers emanate from the origin $\ms{O}$ in a Milne-like spacetime (see Figure \ref{comoving figure in intro}), a possible physical interpretation of this classification would be that these are the particles created at the big bang. 
If this interpretation is correct, then the principal series would correspond to the particles of the standard model (in analogy to Wigner's classification of the Poincar{\'e} group \cite{Wigner_classification}). But this leaves the complementary series up to interpretation. Perhaps
\begin{align*}
\text{complementary series} \,&=\, \text{dark matter particles?} 
\end{align*}

But is there any evidence that dark matter is comprised of spin 0 particles? Yes. Scalar field dark matter (SFDM)  \cite{GuzmanMatos, GuzmanLopez,  MaganaMatos, AbrilRoblesMatos} also known as Bose-Einstein condensate (BEC) dark matter \cite{SeidelSuen, Sin, SinJi, MatosLopez, LopezArgelia, SKM} also known as wave dark matter (WDM) \cite{Bray1, Bray2, GoetzThesis, ParryThesis} also known as fuzzy dark matter (FDM) \cite{FDM1, FDM2} all use the Klein-Gordon wave equation (i.e. the wave equation for spin 0 particles) to model dark matter. The difference in name comes from a difference in motivation. One reason for introducing models of dark matter based on the Klein-Gordon equation is to alleviate the cusp problem associated with the weakly interacting massive particle (WIMP) models of dark matter \cite{LeeDarkMatter}. Furthermore, the models based on the Klein-Gordon equation reproduce the observed spiral pattern density in disk galaxies (see Figures 1 - 4 in \cite{Bray1}) which makes these models promising.


If the identification ``principal series = normal matter" and ``complementary series = dark matter" is true, then the distinguishing feature could be related to the parameter $\nu$ which is determined by Casimir operators built out of rotations and Lorentz boosts (see equation 10.3-1 in \cite{Tung}). Perhaps this could offer an explanation for dark matter's lack of interaction with electromagnetism.

\medskip
\medskip

\noindent{\bf What lies beyond $\tau = 0$?}

\medskip

Recall the Lorentz group $\text{L} = \text{O}(1,3)$ has four connected components $\text{L}^\uparrow_{+}$, $\text{L}^\uparrow_-$, $\text{L}^\downarrow_{+}$, $\text{L}^\downarrow_{-}$. The $\pm$ corresponds to $\det \Lambda = \pm 1$, the $\uparrow$ corresponds to $\Lambda^0_{\:\:\:0} \geq 1$, and the $\downarrow$ corresponds to $\Lambda^0_{\:\:\:0} \leq -1$. Since Milne-like spacetimes are defined for $t > 0$, only the subgroup $\text{L}^\uparrow = \text{L}^\uparrow_+ \cup \text{L}^\uparrow_-$ acts by isometries on Milne-like spacetimes.

What about $\text{L}^\downarrow = \text{L}^\downarrow_+ \cup \text{L}^\downarrow_-$? If $\text{L}^\downarrow$ also acts by isometries, then there would be a universe isometric to ours with the isometry given by the $\text{PT}$ transformation $(t,x,y,z) \mapsto (-t,-x,-y,-z)$. Given the CPT theorem \cite{PCT}, perhaps the universe's missing antimatter is contained in the PT symmetric universe. In section \ref{antimatter section} we show how one can interpret the PT symmetric universe as an antimatter universe via a Lorentz invariant Dirac equation.

We remark this idea is closely related to the same idea in \cite{TBF}. There the authors consider a $k = 0$ FLRW spacetime with metric $g = -d\tau^2 + a^2(\tau)\big[dx^2 + dy^2 + dz^2\big]$ in a radiation dominated era $a(\tau) \propto \sqrt{\tau}$. By moving to conformal time $\tilde{\tau}$ given by $d\tilde{\tau} = d\tau/a(\tau)$, one arrives at the metric $g = a^2(\tilde{\tau})\big[-d\tilde{\tau}^2 + dx^2 + dy^2 + dz^2 \big]$ where $a(\tilde{\tau}) \propto \tilde{\tau}$. They then analytically extend the function $a(\tilde{\tau})$ from $(0, + \infty)$ to $\R$ and call the $(-\infty, 0)$ part the `CPT-symmetric' universe. However, at $\tilde{\tau} = 0$, the metric is $g = 0$. Hence it's degenerate. Therefore this is not a spacetime extension.

\medskip
\medskip

\begin{figure}[h]
\[
\begin{tikzpicture}[scale = 0.5]

\shadedraw [white] (2,2) -- (6,-2) -- (10,2);	
\draw [dashed, thick, blue] (6,-2) -- (2,2);
\draw [dashed, thick, blue] (6,-2) -- (10,2);

\shadedraw[top  color = white, bottom color = gray, thick, white] (2,-6) -- (6,-2) -- (10,-6);
\draw [dashed, thick, blue] (6,-2) -- (2,-6);
\draw [dashed, thick, blue] (6,-2) -- (10,-6);

\draw [<->,thick] (6,-6.5) -- (6,2.5);
\draw [<->,thick] (1.5,-2) -- (10.5,-2);

\draw (6.3550,2.90) node [scale = .85] {$t$};
\draw (10.90, -2.25) node [scale = .85] {$x^i$};


\shade(-13,2) -- (-13,-2) -- (-5,-2) -- (-5,2);

\shade[top  color = white, bottom color = gray](-13,-2) -- (-13,-6) -- (-5,-6) -- (-5,-2);

\draw [<->,thick] (-9,-6.5) -- (-9,2.5);

\draw (-8.5,2.90) node [scale = .85] {$\tilde{\tau}$};

\draw[snake=zigzag, red, thick] (-13,-2) -- (-5,-2);

\draw (-14.5,-2) node [scale = .85] {\small{$\tilde{\tau} = 0$}};


\draw [->] [thick] (3.75,2.8) arc [start angle=30, end angle=0, radius=80pt];

\draw (3.1,3.3) node [scale = .85] {\small{Our universe}};
	
	\draw [->] [thick] (8.25,-6.5) arc [start angle=-150, end angle=-180, radius=80pt];	

\draw (8.75,-7.1) node [scale = .85] {\small{Antimatter universe} };


\draw [->] [thick] (-11.75,2.8) arc [start angle=30, end angle=0, radius=80pt];

\draw (-12,3.3) node [scale = .85] {\small{Our universe}};

\draw [->] [thick] (-6.25,-6.5) arc [start angle=-150, end angle=-180, radius=80pt];	

\draw (-5.75,-7.1) node [scale = .85] {\small{Antimatter universe}};


%
%
%
%
%
%
%
%
	
\end{tikzpicture}
\]
\caption{\small{ 
The figure on the left represents the universe/antimatter universe pair in \cite{TBF}. The metric is degenerate at $\tilde{\tau} = 0$, so the pair together do not form a spacetime. The figure on the right represents the universe/antimatter universe pair for a Milne-like spacetime. In this case the pair coexist in a single nondegenerate spacetime. They are causally connected at the origin $\ms{O}$ where Lorentz invariance holds. 
}}
\end{figure}
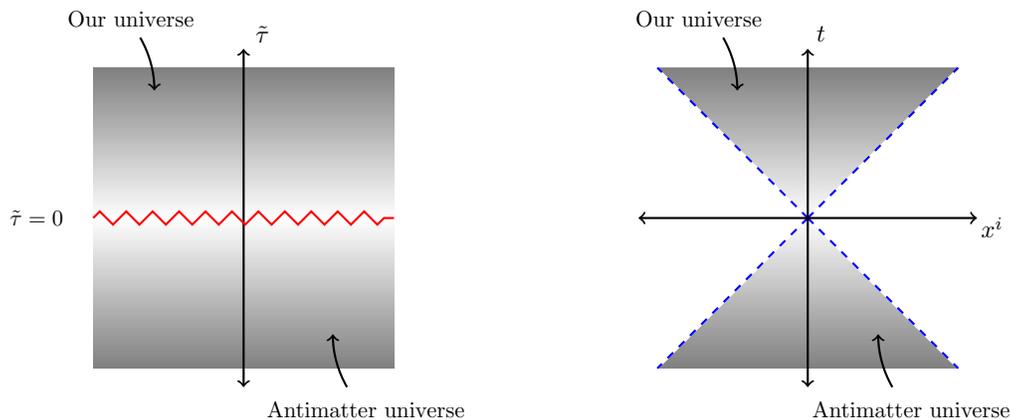

\newpage

\subsection{Open problems}

\medskip

\begin{itemize}

\item[(1)] Is $\tau = 0$ a coordinate singularity for $k =1$ and $k= 0$ FLRW spacetimes?  From \cite{GalLing_con} it is known that no extension can exist with spherical symmetry.

\end{itemize}

\noindent Open problems (2) - (4) refer to the past boundary $\pd^-M$ of a spacetime (Definition \ref{boundary def}).

\begin{itemize}

\item[(2)] Is $\tau = 0$ a coordinate singularity for Milne-like spacetimes with compact $\tau$-slices? The null expansion $\theta$ of the future lightcone  in Minkowski space diverges as one approaches the origin $\ms{O}$ along the cone.
This suggests that, in the compact case, the past boundary $\pd^-M$ cannot be compact (if nonempty).

\medskip

\item[(3)] To understand what can lie beyond $\tau = 0$, it is desired to understand the maximal analytic extension whenever $\Omega$ is analytic on $M \cup \pd^-M$. Minkowski space is the maximal analytic extension of the Milne universe. De Sitter space is the maximal analytic extension of the Milne-like spacetime with scale factor $a(\tau) = \sinh(\tau)$. Therefore we suggest 

\medskip

Conjecture: Let $(M,g)$ be a Milne-like spacetime with an analytic $\Omega$ on $M \cup \pd^-M$. If $(M,g)$ is asymptotically flat (i.e. admits a null scri structure), then the maximal analytic extension contains a noncompact Cauchy surface. If $(M,g)$ is asymptotically de Sitter (i.e. admits a spacelike scri structure), then the maximal analytic extension contains a compact Cauchy surface. A discussion of analytic spacetime extensions and when they are unique can be found in section 4.6 of \cite{ChruCortier}.
\medskip

\item[(4)] Eventually one would want to study models that are not perfectly homogeneous and isotropic. This poses a natural question: what is the correct initial value problem for the lightcone $\pd^-M$ which adequately describes our universe? For example, the null geodesics on $\pd^-M$ emanating from $\ms{O}$ never focus and so there are no null conjugate points on $\pd^-M$. How does this constrain the initial data? Some other ideas/questions for initial conditions on $\pd^-M$ are listed below.

\begin{itemize}

\item[-] Perhaps a constant scalar curvature on $\pd^-M$ should be assumed. This could offer a geometric origin for how the cosmological constant $\Lambda$ appears as an initial condition. 

\medskip

\item[-] Perhaps one should assume the Weyl curvature tensor vanishes on $\pd^-M$ which would be in accordance with Penrose's Weyl curvature hypothesis.

\medskip

\item[-] What initial conditions would adequately describe Lorentz invariance?

\medskip

\item[-] What initial conditions force a rigidity result like the one in section \ref{rigidity section}?

\end{itemize}

\end{itemize}

\section{Definition of Coordinate and Curvature Singularities}\label{definition section}

\subsection{Spacetimes}

Let $k \geq 0$ be an integer. A  $C^k$ \emph{manifold} of dimension $n+1$ is a topological space $M$ endowed with a maximal $C^k$-atlas $\mc{A}$ of dimension $n+1$.  A \emph{coordinate system} is an element $\phi \in \mc{A}$. Specifically, a coordinate system is a $C^{k}$-diffeomorphism $\phi \colon U \to \phi(U) \subset \R^{n+1}$ where $U$ is an open subset of $M$. The coordinate system $\phi$ introduces \emph{coordinates} which are $C^{k}$ maps $x^\mu \colon U \to \R$ via $x^\mu = \pi^\mu \circ \phi$ where $\pi^\mu \colon \R^{n+1} \to \R$ is the canonical projection and $\mu$ runs over the indices $0, 1, \dotsc, n$.  The coordinate systems allow us to define $C^{k}$ curves over $M$. For $k \geq 1$ we use $C^1$ curves to generate tangent vectors at a point $p \in M$. This construction yields the tangent space $T_pM$ and the corresponding $C^{k-1}$ tangent bundle $TM$.

Let $k \geq 0$. A $C^{k}$ \emph{metric} on a $C^{k+1}$ manifold $M$ is a nondegenerate symmetric tensor $g\colon TM \times TM \to \R$ with constant signature whose components $g_{\mu\nu}$ in any coordinate system $\phi \in \mc{A}$ are $C^{k}$ functions. Symmetric means $g(X,Y) = g(Y,X)$ for all $X,Y \in TM$. Nondegenerate means $g(X,Y) = 0$ for all $Y \in TM$ implies $X = 0$. With constant signature means there is an integer $r$ such that at each point $p \in M$, there is a basis $e_0, \dotsc,e_r,\dotsc, e_n \in T_pM$ such that $g(e_\mu, e_\mu) =  -1$ for $0 \leq \mu \leq r$ and $g(e_\mu, e_\mu) = 1$ for $r + 1 \leq \mu \leq n$. If $g(e_0, e_0) = -1 $ and $g(e_i, e_i) = 1$ for all $i = 1, \dotsc, n$, then $g$ is called a \emph{Lorentzian} metric and $(M,g)$ is called a $C^k$ Lorentzian manifold. If $g(e_\mu, e_\mu) = 1$ for all $\mu = 0, 1, \dotsc, n$, then $g$ is called a \emph{Riemannian} metric and $(M,g)$ is called a $C^k$ \linebreak Riemannian manifold. We will only work with Lorentzian manifolds in this paper. Our convention will be that greek indices  $\mu$ and $\nu$ will run through indices $0, 1, \dotsc, n$ and latin indices $i$ and $j$ will run through $1, \dotsc, n$.

Fix $k \geq 0$. Let $(M,g)$ be a $C^k$ Lorentzian manifold. A nonzero vector $X \in T_pM$ is \emph{timelike}, \emph{null}, or \emph{spacelike} if $g(X,X) < 0, \, =0, \,\, > 0$, respectively. A nonzero vector is \emph{causal} if it is either timelike or null.  A Lorentzian manifold $(M,g)$ is \emph{time-oriented} provided there is a $C^0$ timelike vector field $X \in TM$. A causal vector $Y \in T_pM$ is \emph{future-directed} if $g(X,Y) < 0$ and \emph{past-directed} if $g(X,Y) > 0$. Note that $-X$ defines an opposite time-orientation, and so a future-directed causal vector $Y$ with respect to $X$ is a past-directed causal vector with respect to $-X$.

\medskip

\begin{Def}[Spacetime]
\emph{
Let $k \geq 0$. A $C^k$ \emph{spacetime} is a pair $(M,g)$ where $M$ is a connected, Hausdorff, and second-countable $C^{k+1}$ manifold and $g$ is a $C^k$ Lorentzian metric such that $(M,g)$ is time-oriented.
}
\end{Def}

\medskip

\noindent\emph{Remarks:}
\begin{itemize}

\item[-] The Einstein tensor  $R_{\mu\nu} - \frac{1}{2}R g_{\mu\nu}$ requires two derivatives of the metric. Therefore the relevant differentiability class for general relativity is $C^2$.

\item[-] We assume $M$ is connected because we can not make any observations of any other connected components.

\item[-] The Hausdorff condition guarantees uniqueness of limits.

\item[-] The second-countable property allows us to construct partitions of unity whenever needed (e.g. to construct a complete Riemannian metric on $M$). 
\end{itemize}

\medskip
\medskip

Fix $k \geq 0$. Let $(M,g)$ be  a $C^k$ spacetime. A \emph{timelike curve} $\g$ is a piecewise $C^1$ map $\g\colon [a,b] \to M$ such that $\g'(t)$ is future-directed timelike at all its differentiable points, and in the case $t_0 \in [a,b]$ is a break point, we have $\lim_{t \nearrow t_0} \g'(t)$ and $\lim_{t \searrow t_0} \g'(t)$ are both future-directed timelike. If $t_0 = a$ or $ t_0 = b$, then we only require this for the one-sided limit. Note this means that $\g|_{[b-\e,\, b)}$ can be extended to a $C^1$ timelike curve for $\e > 0$ small enough. Letting timelike curves be piecewise $C^1$ allows us to concatenate two timelike curves to form another timelike curve. A \emph{unit} timelike curve is a timelike curve $\g$ such that $g(\g', \g') = -1$ at all its differentiable points. Note that `future-directed' is implicit in our definition of a timelike curve.  
Given a timelike curve $\g \colon [a,b] \to M$, we will often write $\g \subset U$ instead of $\g\big([a,b]\big) \subset U$. Likewise with $\g \cap U$.

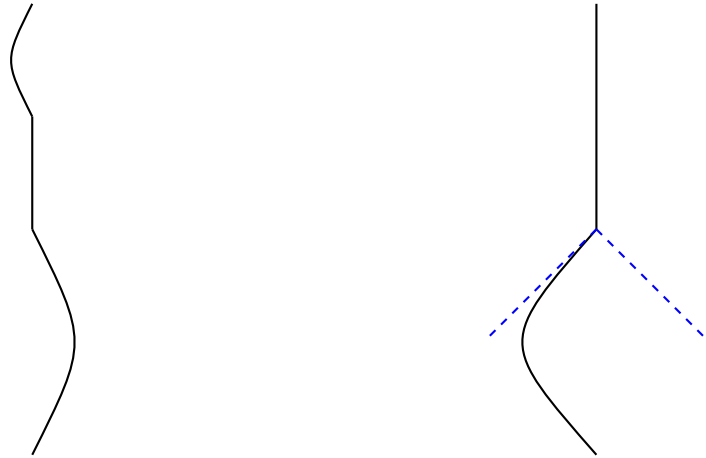
\begin{figure}[h]
\[
\begin{tikzpicture}[scale = 0.75]

\draw [thick, black] (-5,-2) .. controls (-4,0) .. (-5,2);
\draw [thick, black] (-5,2) -- (-5,4);
\draw [thick, black] (-5,4) .. controls (-5.5,5) .. (-5,6);

\draw [thick, black] (5,-2) .. controls (3.25,0) .. (5,2);
\draw[thick, black] (5,2) -- (5,6);

\draw [dashed, thick, blue] (5,2) -- (6.9,0.1);
\draw [dashed, thick, blue] (5,2) -- (3.1,0.1);

\end{tikzpicture}
\]
\captionsetup{format=hang}
\caption{\small{The curve on the left is a timelike curve. The curve on the right is not a timelike curve even though it is timelike at all its differentiable points. We don't consider it a timelike curve because it approaches a null vector at its break point. }}
\end{figure}

Given a point $p \in M$ and an open set $U \subset M$, the \emph{timelike future} of $p$ within $U$, denoted by $I^+(p, U)$, is the set of all points $q \in M$ such that there is a timelike curve $\g \colon [a,b] \to U$ from $p$ to $q$. The \emph{timelike past} of $p$ within $U$, denoted by  $I^-(p, U)$, is defined with the opposite time-orientation. If we wish to emphasize the Lorentzian metric $g$ being used, then we will write $I^+_g(p,U)$. From Proposition 2.6 in \cite{Sbierski}, we know that these are open sets. 

\medskip
\medskip

\begin{prop} \label{timelike is open prop}
$I^+(p, U)$ and $I^-(p,U)$ are open sets.
\end{prop}

\medskip

We reproduce the proof of this proposition in  appendix \ref{I+ open appendix}.

\medskip

\subsection{Spacetime extensions}

\medskip

Coordinate singularities coincide with spacetime extensions. For example, the $r = 2m$ coordinate singularity in Schwarzschild extends the $r > 2m$ region to the maximal analytic extension which contains the $r < 2m$ region. Therefore an understanding of spacetime extensions is needed to understand coordinate singularities. 

Fix $k \geq 0$. Let $(M,g)$ be a $C^{k}$ spacetime. Let $0 \leq l \leq k$. A $C^{l}$ spacetime $(M_\ext, \,g_\ext)$ with the same dimension as $(M,g)$ is a $C^{l}$-\emph{extension} of $(M,g)$ if there is a proper isometric $C^{l+1}$-embedding
\[
(M,g) \:\:\:\: \hookrightarrow \:\:\:\: (M_\ext,\, g_\ext) .
\] 
We identify $M$ with its image under the embedding. The topological boundary of $M$ within $M_\ext$ is denoted by $\pd(M, M_\ext) = \ov{M} \setminus M$.  If $(M_\ext, \, g_\ext)$ is a $C^l$-extension for all $l \geq 0$, then we say $(M_\ext,\, g_\ext)$ is a \emph{smooth} or $C^\infty$-extension.

For the rest of this section, we will fix a $C^l$-extension $(M_\ext, \,g_\ext)$ of a $C^k$ spacetime $(M,g)$.

\medskip
\medskip

\begin{prop}\label{boundaries are nonempty prop}
$\pd(M, M_\ext) \,\neq\, \emptyset$.
\end{prop}

\proof
If this were not true, then $\ov{M} = M$, and so $M_\ext$ would be the disjoint union of the nonempty open sets $M$ and $M_\ext \setminus M$. However, this implies $M_\ext$ is not connected which contradicts the definition of a spacetime.
\qed

\medskip
\medskip

\begin{Def}\label{boundary def}
\emph{Let $(M,g)$ be a $C^k$ spacetime and $(M_\ext, \, g_\ext)$ a $C^l$-extension.
A timelike curve $\g \colon [a,b] \to M_\ext$ is called \emph{future terminating} for a point $p \in \pd (M, M_\ext)$ provided
$\g(b) = p$ and $\g\big([a,b)\big) \subset M$. It is called \emph{past terminating} if $\g(a) = p$ and $\g\big((a,b]\big) \subset M$. 
 The \emph{future} and \emph{past boundaries} of $M$ with respect to $M_\ext$ are
 \medskip
\begin{align*}
\pd^+(M,  M_\ext) \,&=\, \{p \in \pd(M , M_\ext) \mid \text{there is a future terminating timelike curve for } p\}\\\\
\pd^-(M,  M_\ext) \,&=\, \{p \in \pd(M,  M_\ext) \mid \text{there is a past terminating timelike curve for } p\}
\end{align*}
}
\end{Def}

\medskip
\medskip

\noindent\emph{Remark.} If $(M_\ext,\, g_\ext)$ is clear from context, then we will simply write $\pd^+M$ for $\pd^+(M,M_\ext)$. Likewise for $\pd^-M$ and $\pd M$. 

\medskip
\medskip

\begin{figure}[h]
\[
\begin{tikzpicture}[scale = 0.5]

\shadedraw [thick, white](1,-3) -- (5,1) -- (1,5) -- (1,-3);
\draw [dashed, thick, blue] (1, -3) -- (1, 5);
\draw [dashed, thick, blue] (5,1) -- (1, -3);
\draw [dashed, thick, blue] (5,1) -- (1,5);

\draw (2.6,.9) node [scale =.85]{$(M,g)$};
\draw (-2,4) node [scale =.85] {$(M_\ext,\, g_\ext)$};

\draw [->] (-1,0.5) arc [start angle=180, end angle=90, radius=40pt];
\draw (-1,0) node [scale =.75]{$\pd^+M \cap \pd^-M$};

\draw [->] (4.5,4) arc [start angle=0, end angle=-60, radius=40pt];
\draw (4.7,4.5) node [scale =.85] {$\pd^+M$};

\draw [->] (4.5,-2) arc [start angle=0, end angle=60, radius=40pt];
\draw (4.7,-2.5) node [scale =.85] {$\pd^-M$};

\node [scale = .5] [circle, draw, fill = black] at (5,1)  {};
\draw (5.3,0.3) node [scale =.85] {$p$};

\end{tikzpicture}
\]
\captionsetup{format=hang}
\caption{\small{ Various points of $\pd^+M$ and $\pd^-M$ are shown. $p\in \pd M$ is neither in $\pd^+M$ nor in $\pd^-M$.}}
\end{figure}
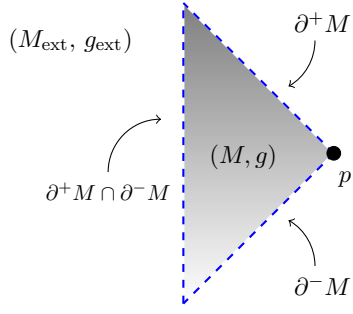

\begin{lem}\label{top lemma}
Let $\g \colon [a,b] \to M_\ext$ be a timelike curve from $p$ to $q$.
\begin{itemize}
\item[\emph{(1)}] If $p \in M$ and $q \notin M$, then $\g$ intersects $\pd^+M$.

\item[\emph{(2)}] If $p \notin M$ and $q \in M$, then $\g$ intersects $\pd^-M$.
\end{itemize}
\end{lem}

\proof
Consider case (1). Define $t_* = \sup \{ t \in [a,b] \mid \g\big([a,t)\big) \subset M\}$. Since $M$ is open we have $t_* > a$. Since $q \notin M$, we have $\g(t_*) \notin M$. On the other hand $\g(t_*)$ is an accumulation point of $M$. Hence $\g(t_*) \in \pd M$. The restriction $\g|_{[a, \, t_*]}$ is a future terminating timelike curve  for $\g(t_*)$. Hence $\g(t_*) \in \pd^+M$. Case (2) follows by reversing the time orientation.
\qed

\medskip 
\medskip

\begin{prop}\label{future or past nonempty cor}
$\pd^+ M \cup \pd^-M \, \neq \, \emptyset$.
\end{prop}

\proof
Fix $p \in \pd M$. Let $U \subset M_\ext$ be an open set around $p$. Fix $q \in I^-(p, U)$. Let $\g \subset U$ be a  timelike curve from $q$ to $p$. We either have $q \in M$ or $q \notin M$. First assume $q \in M$. Then Lemma \ref{top lemma} implies $\g \cap \pd^+M \neq \emptyset$.  Now assume $q \notin M$. Since $p \in \pd M$, the open set $I^+(q, U)$ of $p$ contains a point $r \in M$. Hence there is a timelike curve $\l \subset U$ from $q$ to $r$. Then Lemma \ref{top lemma} shows $\l \cap \pd^-M  \neq \emptyset$.
\qed

%
%

\medskip
\medskip

Proposition \ref{future or past nonempty cor} can be used to show a spacetime $(M,g)$  is \emph{$C^l$-inextendible} (i.e. it admits no $C^l$-extension). Suppose one assumes $(M,g)$ has a $C^l$-extension and then proves $\pd^+M = \emptyset$ and $\pd^-M = \emptyset$, then Proposition \ref{future or past nonempty cor} yields a contradiction. This is how Sbierski proves the $C^0$-inextendibility of Minkowski space and the Schwarzschild spacetime \cite{Sbierski, Sbierski2}.

\medskip
\medskip

\subsection{Definition of coordinate singularities}

\medskip
\medskip

In this section we give a precise definition of what we mean by a `coordinate singularity.' Our goal is to identify when we have made a `poor' choice of coordinates. Before giving the definition, we start with a couple of motivating examples.

\medskip
\medskip

\newpage

\noindent{\bf Motivating examples:}

\begin{itemize}

\item[(1)] Consider
the smooth spacetime $(M,g)$ where $M = (0,\infty) \times \R$ with the metric $g = -\tau^2 d\tau^2 + dx^2$. Since the metric becomes degenerate at $\tau = 0$, we cannot extend $(M,g)$ using the coordinates $(\tau, x)$. However, if we introduce the coordinate $t = \frac{1}{2}\tau^2$, then, with respect to these coordinates, the spacetime manifold becomes $(0, \infty) \times \R$ with metric $-dt^2 + dx^2$. Since the metric is nondegenerate at $t = 0$, we have no problem extending the spacetime using the coordinates $(t,x)$. As such, we say $(\tau, x)$ were a `poor choice' of coordinates, and $\tau = 0$ merely represents a coordinate singularity.
This example demonstrates that a coordinate singularity depends on a spacetime being inextendible within one coordinate system while being extendible in another.

\item[(2)] Consider the smooth spacetime $(M,g)$ where $M = (0,\infty) \times \R$ and $g = -f(t)dt^2 + dx^2$ where $f \colon (0,\infty) \to \R$ is the smooth function given by $f(t) = 1 + \sqrt{t}$. In this case $(M,g)$ extends through $t = 0$ via the spacetime $M_\ext = \R \times \R$ and $g_\ext = -dt^2 + dx^2$ for $t \leq 0$ and $g_\ext = g$ for $t > 0$. However $(M_\ext, g_\ext)$ is not a smooth extension of $(M,g)$. It is only a $C^0$-extension. In this case we would not say that $(t,x)$ are a `poor' choice of coordinates for $(M,g)$, since the coordinates $(t,x)$ can still be used to extend the spacetime just not smoothly.

\end{itemize}

\medskip
\medskip

Fix $k \geq 0$. Let $(M,g)$ be a $C^k$ spacetime with dimension $n+1$. Recall a \emph{coordinate system} is an element of the maximal $C^{k+1}$-atlas for $M$. Specifically, a coordinate system is a $C^{k+1}$-diffeomorphism $\phi \colon U \to \phi(U) \subset \R^{n+1}$ where $U$ is an open subset of $M$. 


Let $\phi \colon U \to \R^{n+1}$ be a coordinate system for a $C^k$ spacetime $(M,g)$. Let $\Omega = \phi(U) \subset \R^{n+1}$. Then $(U,g)$ is $C^{k}$-isometric to $(\Omega,\, \phi_* g)$ where $\phi_*$ is the push forward. Let $0 \leq l \leq k$. Suppose there exists an open set $\Omega' \subset \R^{n+1}$ which properly contains $\Omega$ and a $C^l$-Lorentzian metric $g'$ on $\Omega'$ such that $(\Omega',\, g')$ is a $C^l$-extension of $(\Omega,\, \phi_*g)$. Then we say $(\Omega',\, g')$ is a \emph{$C^l$-coordinate extension} of $(\Omega,\, \phi_* g)$. If such an $(\Omega',\, g')$ exists, then we say $\phi$ is \emph{not $C^l$-maximal}. If no such $(\Omega',\, g')$ exists, then we say $\phi$ is \emph{$C^l$-maximal.} For example, the coordinates $(\tau, x)$ in the first example above are $C^0$-maximal. The coordinates $(t,x)$ in the second example are $C^1$-maximal but not $C^0$-maximal.

\medskip
\medskip

\noindent\emph{Remark.} Intuitively, a coordinate system $\phi$ is $C^0$-maximal if the metric components $g_{\mu\nu}$ with respect to $\phi$ are maximally extended. If they were extended any further, then there would be a degeneracy in the metric (e.g. Schwarzschild).

\medskip
\medskip

\begin{Def}[Coordinate singularity]
\emph{
Fix $k \geq 0$ and $0 \leq l \leq k$. Let $(M,g)$ be a $C^k$ spacetime and let $\phi \colon U \to \R^{n+1}$ be a coordinate system which is $C^0$-maximal. We say $\phi$ admits a \emph{$C^l$-coordinate singularity} for $(M,g)$ if there is a $C^l$-extension $(M_\ext, \, g_\ext)$ and a coordinate system $\psi \colon V \to \R^{n+1}$ for $M_\ext$ such that
\[
V \cap M \,=\, U \:\:\:\:\:\: \text{ and } \:\:\:\:\:\: V \cap (M_\ext \setminus M) \,\neq\, \emptyset.
\]
}
\end{Def}

\medskip
\medskip

\noindent\emph{Remark.} In the definition  $\phi$ represents the `poor' choice of coordinates. $\psi$ represents the `better' choice of coordinates. 

\medskip
\medskip

 For example consider $(M,g)$ from  example (1) above. The coordinate system $\phi = (\tau, x)$ is a $C^\infty$-coordinate singularity for $(M,g)$. This follows because
 
 \begin{itemize}
 \item[-] $\phi$ is $C^0$-maximal
 
 \item[-] $(M_\ext, \, g_\ext)$ is a $C^\infty$-extension of $(M,g)$ where $M_\ext = \R^2$ and $g_\ext = -dt^2 + dx^2$.
 
 \item[-] In this example we simply take $U = M$ and $V = M_\ext$ and $\psi = (t,x)$.
 \end{itemize}

\medskip

\subsection{Definition of curvature singularities}

\medskip
\medskip

In this section we give a precise definition of what we mean by a `curvature singularity.' Before doing so, let's fix notation by recalling the definition of curvature. 

Fix $k \geq 2$. Let $(M,g)$ be a $C^k$ spacetime and $\nabla$ its unique compatible affine connection. Then the \emph{Riemann curvature tensor} is the $(3,1)$ tensor defined by 
\[
R(X,Y)Z \,=\, \nabla_X(\nabla_Y Z) \,-\, \nabla_Y(\nabla_X Z) \,-\, \nabla_{[X,Y]}Z
\]
where $[X,Y]$ is the Lie derivative of $Y$ with respect to $X$. Let $\{\pd_\mu\}$ be a coordinate vector basis with dual one-form basis $\{dx^\mu\}$. The \emph{components} of the Riemann curvature tensor with respect to $\{\pd_\mu\}$ are defined by $R_{\mu\nu\a}^{\:\:\:\:\:\:\:\: \b} = dx^\b \big(R(\pd_\mu, \pd_\nu)\pd_\a\big)$. Using index notation $\nabla_X Y = (X^\mu \nabla_\mu Y^\nu) \pd_\nu $ and the linear and Leibniz properties of the affine connection, we have \[R_{\mu\nu\a}^{\:\:\:\:\:\:\:\: \b}Z^\a = (\nabla_\mu \nabla_\nu - \nabla_\nu\nabla_\mu )Z^\b.\] Here we see the non-commutativity of the second covariant derivatives of $Z$ expressed in terms of the components of the  curvature tensor.

Since gravitation is mathematically described by curvature (e.g. the Einstein equations and tidal acceleration), we can use divergences in the curvature tensor as a way to identify the breakdown of general relativity. Since we want to measure divergences in a coordinate-independent way, we construct curvature invariants out of scalar quantities.   

\medskip
\medskip

\begin{Def}
\emph{
Fix $k \geq 2$. Let $(M,g)$ be a $C^k$ spacetime. A \emph{curvature invariant} on $(M,g)$ is a scalar function which is a polynomial in the components of the metric $g_{\mu\nu}$, its inverse $g^{\mu\nu}$, and the curvature tensor $R_{\mu\nu\a}^{\:\:\:\:\:\:\:\: \b}$.
}
\end{Def}
%
%
%

\medskip
\medskip

\noindent{\bf Examples of curvature invariants:}

\begin{itemize}

\item[(1)] The scalar curvature $R = g^{\mu\nu}R_{\mu\nu} = g^{\mu\nu}R_{\mu\a\nu}^{\:\:\:\:\:\:\:\: \a} $.

\item[(2)] The Kretschmann scalar $R_{\mu\nu\a\b}R^{\mu\nu\a\b}$.

\item[(3)] $R_{\mu\nu}R^{\mu\nu}$.

\end{itemize}

\medskip
\medskip

Fix $k \geq 0$. Let $(M,g)$ be a $C^k$ spacetime. A \emph{future inextendible} timelike curve is a curve $\g \colon [a, b) \to M$  such that for any $a < c < b$, the restriction $\g|_{[a,c]}$ is a timelike curve, and for all $p \in M$ the extended function $\g_p \colon [a,b] \to M$ is not continuous where $\g_p$ is given by $\g_p(t) = \g(t)$ for all $a \leq t < b$ and $\g_p(b) = p$. \emph{Past inextendible} timelike curves are defined time-dually.

\medskip
\medskip

\begin{Def}[Curvature singularity]
\emph{
Fix $k \geq 2$. Let $(M,g)$ be a $C^k$ spacetime. We say $(M,g)$ admits a \emph{future curvature singularity} if there is a future inextendible timelike curve $\g \colon [a,b) \to M$ and a curvature invariant $C$ such that $C \circ \g(t)$ diverges as $t \to b$. Time-dualizing the definition gives \emph{past} curvature  singularities.
}
\end{Def}

\medskip
\medskip

\subsection{A classical example: the Schwarzschild spacetime}

\medskip
\medskip

In this section we apply our definitions of coordinate and curvature singularities to the Schwarzschild spacetime. We will show how the $r = 2m$ event horizon in Schwarzschild is just a coordinate singularity and how $r = 0$ is a curvature singularity. 

\medskip
\medskip

\begin{Def}
\emph{
Let $m > 0$. Define two manifolds 
\begin{align*}
M_\safe \,&=\, \R \times (2m, \infty) \times S^2
\\
M_\unsafe \,&=\, \R \times (0, 2m) \times S^2
\end{align*}
and the metric 
\begin{equation}
g \,=\, -\left(1 - \frac{2m}{r} \right)dt^2 + \left(1 - \frac{2m}{r} \right)^{-1}dr^2 \,+\, r^2 d\Omega^2
\end{equation}
where $(S^2, d\Omega^2)$ is the usual round two-sphere. Then $(M_\safe, \, g)$ and $(M_\unsafe, \, g)$ are the \emph{safe} and \emph{unsafe} Schwarzschild spacetimes.
}
\end{Def}

\medskip
\medskip

Let $0 < \theta < \pi$ and $0 < \phi < 2\pi$ be the standard coordinates on $S^2$. Let $U\subset M_\safe$ be the open set $U = \R \times (2m, \infty) \times (0,\pi) \times (0,2\pi)$. Then the metric in the coordinate system $\xi = (t,r,\theta,\phi)\colon U \to \R^4$ is 
\begin{equation}
g \,=\, -\left(1 - \frac{2m}{r} \right)dt^2 \,+\, \left(1 - \frac{2m}{r} \right)^{-1}dr^2 \,+\, r^2 (d\theta^2 + \sin^2 \theta d\phi^2).
\end{equation}
Note that the coordinate system $\xi \colon U \to \R^4$ is $C^0$-maximal.

Now we introduce other coordinates $(v,r,\theta, \phi)$ given by $v = t +r^*(r)$ where $r^*(r) = r + 2m\log(r/2m - 1)$. The metric in these coordinates is 
\begin{equation}\label{eddington coord eq}
g \,=\, -\left(1 - \frac{2m}{r} \right)dv^2 \,+\, (dv\otimes dr + dr\otimes dv) \,+\, r^2 (d\theta^2 + \sin^2 \theta d\phi^2).
\end{equation}
There is no degeneracy at $r = 2m$ with respect to these coordinates. Therefore we can define a $C^\infty$-extension $(M_\ext,\, g_\ext)$ where $M_\ext = \R\times (0,\infty) \times S^2$ and 
\begin{equation}
g_\ext \,=\, -\left(1 - \frac{2m}{r} \right)dv^2 \,+\, (dv\otimes dr + dr\otimes dv)\,+\, r^2 d\Omega^2.
\end{equation}
Let $V = \R \times (0,\infty) \times (0,\pi) \times (0,2\pi)$. Then $\zeta = (v,r,\theta,\phi) \colon V \to \R^4$ is a coordinate system for $(M_\ext,\, g_\ext)$ such that $V \cap M_\safe = U$ and $V \cap (M_\ext \setminus M_\safe) \neq \emptyset$. Thus

\medskip
\medskip

\begin{prop}
$\xi = (t, r, \theta, \phi)$ admits a $C^\infty$-coordinate singularity for $(M_{\rm safe},\,g)$. 
\end{prop} 

\medskip
\medskip

It's not hard to see that $(M_\unsafe, \,g)$ is $C^\infty$-isometric to the region $r < 2m$ of $(M_\ext,\, g_\ext)$. We end this section by demonstrating that $(M_\unsafe, \, g)$ admits a future curvature singularity.

\medskip
\medskip

\begin{prop}
$(M_{\rm unsafe},\, g)$ admits a future curvature singularity.
\end{prop}

\proof
Consider the future inextendible timelike curve $\g \colon [m, 0) \to M_\unsafe$ given by $\g(r) = (t_0, r, \theta_0, \phi_0)$. Recall that $\pd /\pd r$ is timelike on $M_\unsafe$.  Let $C = R_{\mu\nu\a\b}R^{\mu\nu\a\b}$ denote the Kretschmann scalar. Then a calculation shows $C = 48 m^2/r^6$. Therefore $C \circ \g(r) \to \infty$ as $r \to 0$.
\qed

\section{The Coordinate Singularity for Milne-like Spacetimes}\label{coord singularity section}

\medskip

Let $I \subset \R$ be an open interval. Let $(\S,h)$ be a three-dimensional complete Riemannian manifold with constant sectional curvature. We say $(M,g)$ is an \emph{FLRW spacetime} if $M = I \times \S$ and $g = -d\tau^2 + a^2(\tau)h$
where $a \colon I \to (0,\infty)$ is a continuous function called the \emph{scale factor}.  The integral curves of $\pd / \pd \tau$ are called the \emph{comoving observers}. Physically, they model the trajectories of galaxies.

\medskip
\medskip

\noindent\emph{Remark.} We don't assume any differentiability assumption on the scale factor. Therefore the lowest regularity class for FLRW spacetimes is $C^0$.

\medskip
\medskip

Let $(\R^3, h)$ be hyperbolic space with sectional curvature $k = -1$. Let $(M,g)$ be the corresponding FLRW spacetime. We use the standard coordinates $\xi = (\tau, R, \theta, \phi)$ for $M$ where $\xi \colon U \to \R^4$ and $U = I \times (0,\infty) \times(0,\pi) \times (0, 2\pi)$. With respect to the coordinate system $\xi = (\tau, R, \theta, \phi)$, the metric is
\begin{equation}
g \,= \, -d\tau^2 \,+\, a^2(\tau)\big[dR^2 + \sinh^2(R)(d \theta^2 + \sin^2\theta \, d\phi^2) \big].
\end{equation}

We will first demonstrate how $\xi = (\tau, R, \theta, \phi)$ admits a $C^\infty$-coordinate singularity for $(M,g)$ in two familiar cases: (1) when $(M,g)$ is the Milne universe and (2) when $(M,g)$ is the open-slicing coordinate system for de Sitter space. Then we will show how $\xi = (\tau, R, \theta, \phi)$ admits a $C^0$-coordinate singularity for a class of inflationary spacetimes which we have dubbed `Milne-like.' The justification for calling these `inflationary'  comes in Section \ref{horizon problem section} where we show that these spacetimes solve the horizon problem in cosmology.

In section \ref{C2 section} we show that Milne-like spacetimes do not admit curvature singularities provided the scale factor satisfies $a''(\tau) = \a\tau + C \tau^3 + o(\tau^3)$ with $\a, C \in \R$. In section \ref{rigidity section} we show a rigidity result: if a Milne-like spacetime satisfies both the weak and strong energy conditions, then it must be the Milne universe.

\medskip
\medskip

\subsection{The Milne universe}

\medskip

Let $(\R^3, h)$ be hyperbolic space with sectional curvature $k = -1$. The Milne universe is the corresponding FLRW spacetime $(M,g)$ given by $M = (0, \infty) \times \R^3$ and with scale factor $a(\tau) = \tau$. With respect to the coordinate system $\xi = (\tau, R, \theta, \phi)$, the metric is
\begin{equation}
 g\,=\, -d\tau^2 \,+\, \tau^2\big[dR^2 + \sinh^2(R)\,(d \theta^2 + \sin^2\theta \, d\phi^2) \big].
\end{equation}
We introduce a new coordinate system $\zeta = (t,r,\theta, \phi)$ where $\theta$ and $\phi$ are unchanged, but $t$ and $r$ are given by
\begin{equation}
t \,=\, \tau \cosh(R) \:\:\:\: \text{ and } \:\:\:\: r \,= \,\tau \sinh(R).
\end{equation}
Then we have $-dt^2 + dr^2 = -d\tau^2 + \tau^2 dR^2$, so that  the metric in the coordinate system $\zeta = (t,r, \theta, \phi)$ is 
\begin{equation}
g \,=\, -dt^2 + dr^2 + r^2(d\theta^2 + \sin^2 \theta \, d\phi^2)
\end{equation}
which is just the usual Minkowski metric. The coordinate system $\xi = (\tau, R, \theta, \phi)$ is $C^0$-maximal, but we can find a $C^\infty$-extension via $\zeta = (t,r, \theta, \phi)$. Therefore

\medskip
\medskip

\begin{prop}
$\xi = (\tau, R, \theta, \phi)$ admits a $C^\infty$-coordinate singularity for $(M,g)$.
\end{prop}

\medskip
\medskip

 The constant $\tau$ slices are hyperboloids sitting inside the future lightcone of the origin. We take the extension to be $(M_\ext,\, g_\ext) =$ Minkowski space. As $\tau \to 0$, these slices approach the lightcone at the origin $\ms{O}$ in Minkowski space where the extended metric $g_\ext$ is nondegenerate.

\begin{figure}[h]
\[
\begin{tikzpicture}[scale = .75]

\shadedraw [white] (-4.1,2.1) -- (0,-2) -- (4.1,2.1);
\draw [dashed, thick, blue] (0,-2) -- (4.1,2.1);
\draw [dashed, thick, blue] (0,-2) -- (-4.1,2.1);

\draw [<->,thick] (0,-3.5) -- (0,2.35);
\draw [<->,thick] (-4.5,-2) -- (4.5,-2);

\draw (-.35,2.5) node [scale = .85] {$t$};
\draw (4.75, -2.25) node [scale = .85] {$x^i$};
\draw (-.25,-2.25) node [scale = .85] {$\ms{O}$};

\draw (-2,-3) node [scale = .85] {\small{$(M_\ext,\, g_\ext)$}};

\draw [->] [thick] (1.5,2.8) arc [start angle=140, end angle=180, radius=60pt];
\draw (2.0,3.25) node [scale = .85]{\small{The Milne universe}};

\draw [->] [thick] (-2.4,-1.8) arc [start angle=-90, end angle=-30, radius=40pt];
\draw (-3,-1.7) node [scale = .85] {\small $\pd^-M$};


\draw [thick, red] (-3.84,2.1) .. controls (0,-2) .. (3.84,2.1);
\draw [thick, red] (-3.5,2.1) .. controls (0, -1.3).. (3.5,2.1);

\draw [->] [thick]  (1,-2.3) arc [start angle=-120, end angle=-180, radius=40pt];
\draw (2.3,-2.5) node [scale = .85] {\small{$\tau =$ constant }};

\end{tikzpicture}
\]
\captionsetup{format=hang}
\caption{\small{The Milne universe sits inside the future lightcone of the origin $\ms{O}$ in the extension which is just Minkowski space. It's foliated by constant $\tau$ slices which are hyperboloids.  }}\label{milne-like figure}
\end{figure}
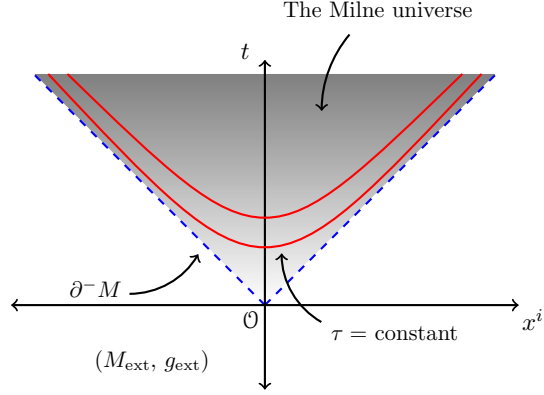

\medskip
\medskip

\subsection{De Sitter space}

\medskip

The \emph{open slicing coordinate system} for de Sitter space is a $k = -1$ FLRW spacetime $M = (0, \infty) \times \R^3$ with scale factor $a(\tau) = \sinh(\tau)$. With respect to the coordinate system $\xi = (\tau, R, \theta, \phi)$, the metric is 
\begin{equation}
g\, = \, -d\tau^2 + \sinh^2(\tau)\big[dR^2 + \sinh^2(R)\, (d\theta^2 + \sin^2\theta \, d\phi^2 )\big].
\end{equation}
We introduce a new coordinate system $\zeta = (t,r, \theta, \phi)$ where $\theta$ and $\phi$ are unchanged, but $t$ and $r$ are given by 
\begin{equation}
t\, = \, b(\tau) \cosh(R) \:\:\:\: \text{ and } \:\:\:\: r \,= \, b(\tau) \sinh(R),
\end{equation}
where $b(\tau) = \tanh(\tau/2) = \sinh \tau/ (1 + \cosh \tau)$. Then $b'(\tau) = b(\tau)/a(\tau)$, and so we have the following relationship between $(t, r)$ and $(\tau, R)$. 
\begin{equation}
\left(\frac{a(\tau)}{b(\tau)}\right)^2\big( -dt^2 + dr^2\big) \, =  \, -d\tau^2 + a^2(\tau) dR^2.
\end{equation}
Therefore the metric is
\begin{equation}
g \, = \, \left(\frac{a(\tau)}{b(\tau)}\right)^2\big[-dt^2 + dr^2 + r^2(d\theta^2 + \sin^2 \theta \,d\phi^2)\big],
\end{equation}
which is conformal to the Minkowski metric. 
 Using $b(\tau) = \tanh(\tau/2)$ and $b^2(\tau) = t^2 - r^2$, we have $\tau = 2 \tanh^{-1}(\sqrt{t^2 - r^2})$. Therefore
$1/b'(\tau) = a(\tau)/b(\tau) =  2/(1 - t^2 + r^2)$.  Thus the metric in the coordinate system $\zeta = (t, r, \theta, \phi)$  is
\begin{equation}
g \,=\, \left(\frac{2}{1 - t^2 + r^2}\right)^2\big[-dt^2 + dr^2 + r^2(d\theta^2 + \sin^2 \theta \, d\phi^2)\big].
\end{equation}

The coordinate system $\xi = (\tau, R, \theta, \phi)$ is $C^0$-maximal, but  we can define a $C^\infty$-extension via $\zeta = (t,r,\theta, \phi)$.  Thus

\medskip
\medskip

\begin{prop}
$\xi = (\tau, R, \theta, \phi)$ admits a $C^\infty$-coordinate singularity for $(M,g)$.
\end{prop}

\medskip
\medskip

The constant $\tau$ slices are hyperboloids sitting inside the future lightcone at the origin. We take the extension to be $(M_\ext,\, g_\ext) =$ a smooth spacetime conformal to Minkowski space. As $\tau \to 0$, these slices approach the lightcone where the extended metric $g_\ext$ is nondegenerate. 

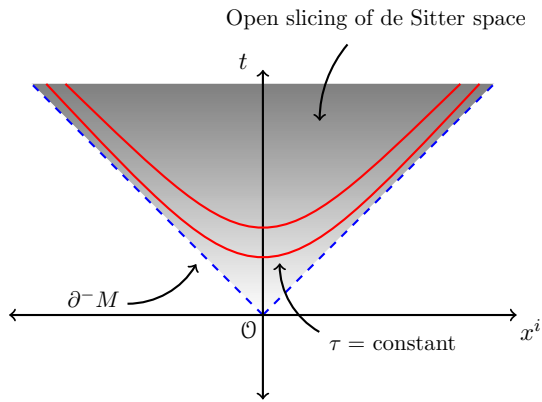
\begin{figure}[h]
\[
\begin{tikzpicture}[scale = 0.75]

\shadedraw [white] (-4.1,2.1) -- (0,-2) -- (4.1,2.1);
\draw [dashed, thick, blue] (0,-2) -- (4.1,2.1);
\draw [dashed, thick, blue] (0,-2) -- (-4.1,2.1);

\draw [<->,thick] (0,-3.5) -- (0,2.35);
\draw [<->,thick] (-4.5,-2) -- (4.5,-2);

\draw (-.35,2.5) node [scale = .85] {$t$};
\draw (4.75, -2.25) node [scale = .85] {$x^i$};
\draw (-.25,-2.25) node [scale = .85] {$\ms{O}$};


\draw [->] [thick] (1.5,2.8) arc [start angle=140, end angle=180, radius=60pt];
\draw (2.0,3.25) node [scale = .85]{\small{Open slicing of de Sitter space}};

\draw [->] [thick] (-2.4,-1.8) arc [start angle=-90, end angle=-30, radius=40pt];
\draw (-3,-1.7) node [scale = .85] {\small $\pd^-M$};


\draw [thick, red] (-3.84,2.1) .. controls (0,-2) .. (3.84,2.1);
\draw [thick, red] (-3.5,2.1) .. controls (0, -1.3).. (3.5,2.1);

\draw [->] [thick]  (1,-2.3) arc [start angle=-120, end angle=-180, radius=40pt];
\draw (2.3,-2.5) node [scale = .85] {\small{$\tau =$ constant }};

\end{tikzpicture}
\]
\captionsetup{format=hang}
\caption{\small{The open slicing coordinates of de Sitter space sits inside the future lightcone at the origin $\ms{O}$ in a spacetime conformal to Minkowski space. }}\label{milne-like figure}
\end{figure}

\subsection{Milne-like spacetimes}\label{inflation definition}\label{milne-like spacetimes section}

\medskip

Now we wish to  show that $\xi = (\tau, R, \theta, \phi)$ is a coordinate singularity for scale factors that can model the dynamics of our universe. That is, we wish to show $\tau = 0$ is a coordinate singularity for suitably chosen scale factors $a(\tau)$ which 

\begin{itemize}

\item[-] begin inflationary $a(\tau) \sim \sinh(\tau)$

\item[-] then transitions to a radiation dominated era $a(\tau) \sim \sqrt{\tau}$

\item[-] then transitions to a matter dominated era $a(\tau) \sim \tau^{2/3}$

\item[-] and ends in a dark energy dominated era $a(\tau) \sim e^{\sqrt{\frac{\Lambda}{3}} \tau}$ 

\end{itemize}

\medskip
\medskip

If we assume for small $\tau$, the scale factor satisfies $a(\tau) \sim \tau$, then, by curve fitting, we can use $a(\tau)$ to represent each of the above eras, thus modeling the dynamics of our universe. To make this precise, we assume for small $\tau$, the scale factor satisfies $a(\tau) = \tau + o(\tau^{1+\e})$ for some $\e > 0$ (i.e. $\big[a(\tau) -\tau\big]/\tau^{1+\e} \to 0$ as $\tau \to 0$). In particular any convergent Taylor expansion $a(\tau) = \sum_{n = 1}^\infty c_n\tau^n$ (with $c_1 = 1$) will satisfy this condition.

\medskip
\medskip

\begin{Def}
\text{}
\begin{itemize}
\item[\emph{(1)}] \emph{Let $(M,g)$ be an FLRW spacetime. We say $(M,g)$ is \emph{inflationary} if the scale factor for small $\tau$ satisfies 
$a(\tau) = \tau + o(\tau^{1+\e})$ for some $\e > 0$.}

\item[\emph{(2)}] \emph{We say $(M,g)$ is \emph{Milne-like} if it is an inflationary FLRW spacetime such that $(\S, h) = (\R^3, h)$ where $h$ is the hyperbolic metric with sectional curvature $k = -1$. We assume the coordinate system $\xi = (\tau, R, \theta, \phi)$ is $C^0$-maximal.}
\end{itemize}
\end{Def}

\medskip
\medskip

\noindent\emph{Remarks.} 

\begin{itemize}

\item[-] The motivation for the word `inflationary' comes in Section \ref{horizon problem section} where we show that the particle horizon is infinite for scale factors which obey $a(\tau) = \tau + o(\tau^{1 + \e})$. 

\item[-] A $C^k$ Milne-like spacetime is one such that the spacetime is $C^k$ (i.e. the scale factor $a(\tau)$ is a $C^k$ function).

\item[-] For inflationary spacetimes we have $a(0) := \lim_{\tau \to 0}\, a(\tau) = 0$. 

\item[-] The coordinate system $\xi = (\tau, R, \theta, \phi)$ is defined for all $\tau \in I = (0, \tau_{\rm \max})$ where $\tau_{\rm max} \in (0, +\infty]$. For our universe, we expect $\tau_{\rm max} = +\infty$ due to dark energy.

\end{itemize}

\medskip
\medskip

The next theorem improves and refines Theorem 3.4 in \cite{GalLing_con}.

\medskip
\medskip

\begin{thm}\label{extension thm}
$\xi = (\tau, R, \theta, \phi)$ admits a $C^0$-coordinate singularity for Milne-like spacetimes.
\end{thm}

\proof
Let $(M,g)$ be a Milne-like spacetime. 
With respect to the coordinate system $\xi = (\tau, R, \theta, \phi)$, the metric is 
\begin{equation}
g \, = \,  -d\tau^2 \,+\, a^2(\tau)\big[dR^2 + \sinh^2(R)(d\theta^2 + \sin^2\theta \, d\phi^2) \big].
\end{equation}
Fix any $\tau_0  \in I$. The specific choice does not matter; any $\tau_0 $ will do. Define a new coordinate system $ \zeta = (t,r,\theta, \phi)$ by

\begin{equation}\label{t and r}
t \,=\, b(\tau)\cosh(R ) \:\:\:\: \text{ and } \:\:\:\: r \,=\, b(\tau)\sinh(R)
\end{equation}
where $b \colon I \to (0,\infty)$ is given by
\begin{equation}
b(\tau) \,=\, \exp\left(\int_{\tau_0}^\tau \frac{1}{a(s)}ds \right) .
\end{equation}
Note that $b(\tau)$ is an increasing $C^1$ function and hence it's invertible. Therefore $\tau$ as a function of $t$ and $r$ is 
\begin{equation}\label{def for tau}
\tau \,=\, b^{-1}\big( \sqrt{t^2 - r^2}\big).
\end{equation}
Note that $t$ and $r$ are defined for all points such that $t^2 - r^2 < b^2(\tau_{\rm max})$.  With respect to the coordinate system $\zeta = (t, r, \theta, \phi)$, the metric takes the form 
\begin{align}\label{metric eq}
g \,&=\, \Omega^2\big(\tau(t,r)\big)\big[-dt^2 + dr^2 + r^2(d\theta^2 + \sin^2\theta d\phi^2)\big] 
\end{align}
where
\begin{equation}
\Omega(\tau) \,=\, \frac{1}{b'(\tau)} \,=\, \frac{a(\tau)}{b(\tau)}.
\end{equation}

 Now we prove $\xi = (\tau, R, \theta, \phi)$ admits a $C^0$-coordinate singularity for $(M,g)$.  For this it suffices to show $\Omega(0) := \lim_{\tau \to 0}\Omega(\tau)$ exists and is a finite positive number. Indeed this will imply
the Lorentzian metric given by equation (\ref{metric eq}) extends continuously through $\tau = 0$ which  corresponds to the lightcone $t = r$, i.e. this will imply that $(M,g)$ is $C^0$-extendible via $\zeta = (t,r,\theta,\phi)$. 

To show $0 < \Omega(0) < \infty$, put $b'(0) := \lim_{\tau \to 0}b'(\tau) = \lim_{\tau \to 0} b(\tau)/a(\tau)$. By our definition of an inflationary spacetime, there is an $\e_0 > 0$ such that $a(\tau) = \tau + o (\tau^{1 + \e_0})$. Therefore $\lim_{\tau \to 0}f(\tau)/\tau^{1 + \e_0} = 0$ where $f(\tau)$ is given by $a(\tau) = \tau + f(\tau)$. Therefore for any $\e > 0$, there exists a $\delta > 0$ such that for all $0 < \tau < \delta$, we have $|f(\tau)| < \e\tau^{1 + \e_0}$. Choosing $\e = 1$, we have $\tau - \tau^{1 + \e_0} < \tau + f(\tau) < \tau + \tau^{1 + \e_0}$. Thus $b(\tau)/a(\tau)$ is squeezed between 
\begin{equation}
\frac{1}{a(\tau)}\exp\left(-\int_{\tau}^{\tau_0}\frac{1}{(\tau - \tau^{1 + \e_0})}ds \right) \, < \, \frac{b(\tau)}{a(\tau)} \, < \, \frac{1}{a(\tau)}\exp\left(-\int_\tau^{\tau_0}\frac{1}{(\tau + \tau^{1 + \e_0})}ds \right)
\end{equation}
Evaluating the integrals, we find 
\begin{equation}
\frac{1}{\tau_0}\left(\frac{\tau}{a(\tau)} \right)\left( \frac{1 - \tau^{\e_0}}{1 + \tau_0^{\e_0}}\right)^{-1/\e_0} \, < \, \frac{b(\tau)}{a(\tau)} \, < \, \frac{1}{\tau_0}\left(\frac{\tau}{a(\tau)} \right)\left( \frac{1 + \tau^{\e_0}}{1 + \tau_0^{\e_0}}\right)^{-1/\e_0}
\end{equation}
Since this holds for all $0 < \tau < \delta$, we have $\Omega(0) = 1/b'(0) = \tau_0$. \qed

\medskip
\medskip

\begin{figure}[h]
\[
\begin{tikzpicture}[scale = .85]

\shadedraw [white] (-4.1,2.1) -- (0,-2) -- (4.1,2.1);
\draw [dashed, thick, blue] (0,-2) -- (4.1,2.1);
\draw [dashed, thick, blue] (0,-2) -- (-4.1,2.1);

\draw [<->,thick] (0,-3.5) -- (0,2.35);
\draw [<->,thick] (-4.5,-2) -- (4.5,-2);

\draw (-.35,2.5) node [scale = .85] {$t$};
\draw (4.75, -2.25) node [scale = .85] {$x^i$};
\draw (-.25,-2.25) node [scale = .85] {$\ms{O}$};

\draw (-2,-3) node [scale = .85] {\small{$(M_\ext,\, g_\ext)$}};

\draw [->] [thick] (1.5,2.8) arc [start angle=140, end angle=180, radius=60pt];
\draw (1.75,3.25) node [scale = .85]{\small{A Milne-like spacetime}};

\draw [->] [thick] (-2.4,-1.8) arc [start angle=-90, end angle=-30, radius=40pt];
\draw (-3,-1.7) node [scale = .85] {\small $\pd^-M$};


\draw [thick, red] (-3.84,2.1) .. controls (0,-2) .. (3.84,2.1);
\draw [thick, red] (-3.5,2.1) .. controls (0, -1.3).. (3.5,2.1);

\draw [->] [thick]  (1,-2.3) arc [start angle=-120, end angle=-180, radius=40pt];
\draw (2.3,-2.5) node [scale = .85] {\small{$\tau =$ constant }};

\end{tikzpicture}
\]
\captionsetup{format=hang}
\caption{\small{A Milne-like spacetime sits inside the future lightcone at the origin $\ms{O}$ in a spacetime extension $(M_\ext,\, g_\ext)$ which is conformal to Minkowski space.   }}\label{milne-like figure}
\end{figure}
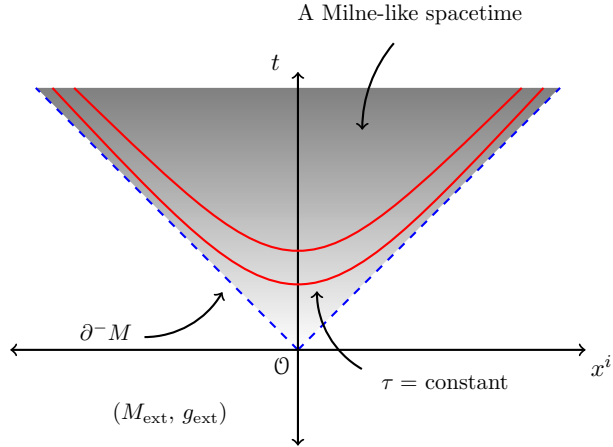

\medskip
\medskip

\noindent\emph{Remark.} The past boundary $\pd^-M$ acts as a Cauchy horizon for the Milne-like spacetime. As such the white region beyond $\pd^-M$ cannot be uniquely determined. 

\newpage

\subsection{Milne-like spacetimes without curvature singularities}\label{C2 section}

\medskip

The scalar curvature for a $k = -1$ FLRW spacetime is given by 

\[
R(\tau) \,=\, \frac{6}{a^2(\tau)}\big[a(\tau)a''(\tau) + a'(\tau)^2 - 1 \big].
\]
For a Milne-like spacetime, we have $a'(0) = 1$.  If the second derivative of the scale factor satisfies $a''(\tau) = \a\tau + o(\tau)$, then we have $a'(\tau) = 1 + \frac{1}{2}\a \tau^2 + o(\tau^2)$. Plugging these into $R(\tau)$, we find $\lim_{\tau \to 0}R(\tau) \,=\, 12\a$. Thus the scalar curvature does not diverge at the big bang.

In this section we generalize the above result and show that Milne-like spacetimes do not admit any curvature singularities provided the second derivative of the scale factor satisfies 
\[a''(\tau) \,=\, \a\tau + C\tau^3 + o(\tau^3)\]
 (i.e. $[a''(\tau) - \a \tau + C\tau^3]/\tau^3 \to 0$ as $\tau \to 0$). Here $\a,C \in \R$ are constants. The limiting condition implies $\a = a'''(0)$.  This limiting condition applies to any convergent Taylor expansion $a(\tau) = \sum_{n = 1}^\infty c_n \tau^n$ with $c_1 = 1$ and $c_2 = 0$ and $c_4 = 0$. Indeed this limiting condition applies to $a(\tau) = \sinh(\tau)$ (i.e. the open-slicing of de Sitter space). This agrees with the fact that de Sitter space has no curvature singularities. 

An example of a Milne-like spacetime where we do have a curvature singularity is given by the scale factor $a(\tau) = \tau + \tau^2$. The scalar curvature diverges as $\tau \to 0$. Indeed in this case we have $a''(\tau) = 2 \neq \a\tau + C\tau^3 + o(\tau^3)$.

%

\medskip
\medskip


\begin{lem}\label{curvature sing lem}
Fix $k \geq 2$. Let $(M,g)$ be a $C^k$ Milne-like spacetime. Suppose the second derivative of the scale factor satisfies $a''(\tau) = \a\tau + C\tau^3 + o(\tau^3)$ where $\a,C \in \R$. Then for any $p \in \pd^-M$, the limits of \[
\frac{\pd \Omega}{\pd t}, \:\:\:\: \frac{\pd \Omega}{\pd r}, \:\:\:\: \frac{\pd^2 \Omega}{\pd t^2}, \:\:\:\: \frac{\pd^2 \Omega}{\pd r^2}
\]
as $(t,r,\theta,\phi) \to p$ all exist and are finite.
\end{lem}

\medskip
\medskip

The proof of Lemma \ref{curvature sing lem} is in appendix \ref{C2 appendix}.

\medskip
\medskip

\begin{thm}
Fix $k \geq 2$. Let $(M,g)$ be a $C^k$ Milne-like spacetime. Suppose the second derivative of the scale factor satisfies $a''(\tau) = \a \tau + C\tau^3 + o(\tau^3)$ where $\a,C \in \R$. Then $(M,g)$ admits no past curvature singularities.
\end{thm}

\proof

\begin{figure}[h]
\[
\begin{tikzpicture}[scale = .85]

\shadedraw [dashed, thick, blue](-4,2) -- (0,-2) -- (4,2);

\draw [<->,thick] (0,-3.5) -- (0,2.25);
\draw [<->,thick] (-4.5,-2) -- (4.5,-2);

\draw (-.35,2.5) node [scale = .85] {$t$};
\draw (4.75, -2.25) node [scale = .85] {$x^i$};
\draw (-.25,-2.25) node [scale = .85] {$\ms{O}$};

\draw (-2,-3) node [scale = .85] {\small{$(M_\ext,\, g_\ext)$}};

\draw [ thick, black] (1,-1) .. controls (0.75,0.5) .. (1.25,1.5);

\draw (.5,.25) node  {$\g$};
 
 \node [scale = .5] [circle, draw, fill = black] at (1,-1)  {};
\draw (1.35,-1.35) node [scale =.85] {$p$};



\end{tikzpicture}
\]
\captionsetup{format=hang}
\caption{\small{A past-inextendible timelike curve $\g$ inside a Milne-like spacetime $(M,g)$ terminates at a past endpoint $p \in \pd^-M$. If $a''(\tau) = \a\tau + C\tau^3 + o(\tau^3)$, then any curvature invariant along $\g$ will limit to a well-defined finite value at $p$.}}\label{curvature singularity figure}
\end{figure}
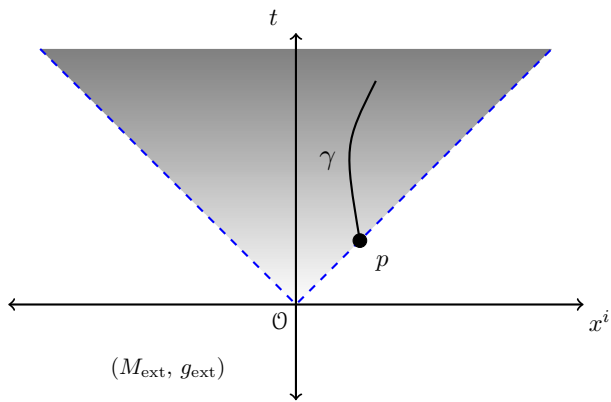

 Let $\g\colon (0, b] \to M$ be any past-inextendible timelike curve parameterized by $\tau$ (we can parameterize by $\tau$ since it's a time function). Since $\g$ is past inextendible and timelike, Figure \ref{curvature singularity figure} shows that there exists a point $p \in \pd^-M$ such that $p = \lim_{\tau \searrow 0} \g(\tau)$. More rigorously, the point $p$ can be determined by writing out $\g$ in the $\zeta = (t, r, \theta, \phi)$ coordinate system.  
 \begin{align*}
t(p) \,&=\, \lim_{\tau \to 0}\, t \circ \g(\tau)
&&r(p)\,=\, \lim_{\tau \to 0}\, r \circ \g(\tau)
\\
\theta(p) \,&=\, \lim_{\tau \to 0}\, \theta \circ \g(\tau)
\,&&\phi(p) =\, \lim_{\tau \to 0}\, \phi \circ \g(\tau) 
 \end{align*}
 The existence of these limits follows from $\g$ being past-inextendible and timelike. Since any curvature invariant is constructed out of first and second derivatives of the metric coefficients (i.e. the first and second derivatives of $\Omega$ in this case), Lemma \ref{curvature sing lem} implies any curvature invariant has a finite-value quantity at $p$. Thus there are no past curvature singularities for $(M,g)$.
\qed

\medskip
\medskip
\medskip
\medskip

\medskip
\medskip

\subsection{A rigidity result}\label{rigidity section}

In this section we will establish the following rigidity result.

\medskip
\medskip

\begin{thm}\label{rigidity thm}
Fix $k \geq 2$. Suppose $(M,g)$ is a $C^k$ Milne-like spacetime which satisfies both the weak and strong energy conditions. Then $(M,g)$ is the Milne universe.
\end{thm}

\medskip
\medskip

Before proving Theorem \ref{rigidity thm}, we recall the definitions of the weak and strong energy conditions.

\medskip
\medskip


\begin{Def}{\rm Fix $k \geq 2$ and let $(M,g)$ be a $C^k$ spacetime.  The Einstein tensor is $G_{\mu\nu} = R_{\mu\nu} - \frac{1}{2}R g_{\mu\nu}$. We say $(M,g)$ satisfies 

\begin{enumerate}

	\item[-] the  \emph{weak energy condition} if $G_{\mu\nu}X^\mu X^\nu \geq 0$ for all timelike $X$.

	\item[-] the \emph{strong energy condition} if $R_{\mu\nu}X^\mu X^\nu \geq 0$ for all timelike $X$.

\end{enumerate}
}
\end{Def}

\medskip
\medskip

Let $(M,g)$ be an FLRW spacetime. Following \cite{ON}, we define the \emph{energy density} $\rho$ and \emph{pressure function} $p$ in terms of the Einstein tensor. If $u = \pd / \pd \tau$ and $e$ is any unit spacelike vector orthogonal to $u$ (its choice does not matter by isotropy), then
\[
\rho \,=\, \frac{1}{8\pi} G_{\mu\nu} u^\mu u^\nu
\:\:\:\:\:\: \text{ and } \:\:\:\:\:\:
p \,=\, \frac{1}{8\pi} G_{\mu\nu}e^\mu e^\nu \]
 We make use of the following proposition. See also equations (9.2.19) and (9.2.20) in \cite{Wald}. 

\medskip
\medskip

\begin{prop}\label{EC prop}
Let $(M,g)$ be an FLRW spacetime. 
\begin{enumerate}

	\item[\emph{(a)}] The weak energy condition is equivalent to $\rho \geq 0$ and $\rho + p \geq 0$.

	\item[\emph{(b)}] The strong energy condition is equivalent to $\rho + 3p \geq 0$ and $\rho + p \geq 0$. 
	
\end{enumerate}

\end{prop}

\medskip
\medskip

The proof of Proposition \ref{EC prop} is in appendix \ref{energy equiv appendix}.

\medskip
\medskip

\noindent \underline{\emph{Proof of Theorem \emph{\ref{rigidity thm}}}}:

\proof
Friedmann's equations are (see equations (5.2.14) and (5.2.15) in \cite{Wald}):
\begin{align}
8\pi \rho \,&=\, \frac{3}{a^2}\big[(a')^2 -1\big] \label{Friedmann eq 1}
\\
8\pi(\rho + 3p) \,&=\, -6\frac{a''}{a} \label{Friedmann eq 2}
\end{align}
The weak energy condition implies $\rho \geq 0$. Therefore $a'(\tau) \geq 1$ for all $\tau$ by equation (\ref{Friedmann eq 1}). The strong energy condition implies $\rho + 3p \geq 0$. Therefore $a''(\tau) \leq 0$ for all $\tau$ by equation (\ref{Friedmann eq 2}). Hence $a'$ is decreasing. Since $a'(0) := \lim_{\tau \to 0} a'(\tau) = 1$, we have $a'(\tau) \leq 1$ for all $\tau$. Therefore $a'(\tau) = 1$ identically, and so $a(\tau) = \tau$ for all $\tau$. Thus $(M,g)$ is the Milne universe.
\qed

\medskip
\medskip

\section{Cosmological Properties of Milne-like Spacetimes}

\medskip
\medskip

\subsection{The geometric solution to the horizon problem}\label{horizon problem section}

\medskip

Our definition for an inflationary FLRW spacetime was one whose scale factor satisfies $a(\tau) = \tau + o(\tau^{1+\e})$ for some $\e > 0$. Our motivation is that these spacetimes solve the horizon problem, and this is true for $k = +1$, $0$, or $-1$. However, what's unique about Milne-like spacetimes is that they extend into a larger spacetime because the big bang is just a coordinate singularity. This offers a new geometrical picture of how Milne-like spacetimes solve the horizon problem as we discuss below. 

We briefly recall the horizon problem in cosmology. It is  the main motivating reason for inflationary theory \cite{WeinbergCos}. The problem comes from the uniform temperature of the CMB radiation. From any direction in the sky, we observe the CMB temperature as 2.7 K. The uniformity of this temperature is puzzling: if we assume the universe exists in a radiation dominated era all the way down to the big bang (i.e. no inflation), then the points $p$ and $q$ on the surface of last scattering don't have intersecting past lightcones. So  how can the CMB temperature be so uniform if $p$ and $q$ were never in causal contact in the past?

\medskip
\medskip

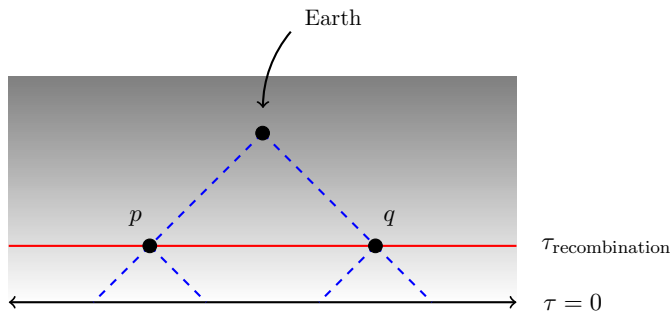
\begin{figure}[h]
\[
\begin{tikzpicture}[scale = 0.75]

\shade(-4.5,2) -- (-4.5,-2) -- (4.5,-2) -- (4.5,2);

\draw [<->,thick] (-4.5,-2) -- (4.5,-2);

\draw [->] [thick] (.5,2.8) arc [start angle=140, end angle=180, radius=60pt];
\draw (1.25,3.05) node [scale = .85]{\small{Earth}};

\draw (5.5, -2) node [scale = .85] {$\tau = 0$};

\draw [thick, red] (-4.5,-1) -- (4.5,-1);

\draw [dashed, thick, blue] (0,1) -- (2,-1);
\draw [dashed, thick, blue] (0,1) -- (-2,-1);

\draw [dashed, thick, blue] (-2,-1) -- (-1,-2);
\draw [dashed, thick, blue] (-2,-1) -- (-3,-2);

\draw [dashed, thick, blue] (2,-1) -- (3,-2);
\draw [dashed, thick, blue] (2,-1) -- (1,-2);

\node [scale = .50] [circle, draw, fill = black] at (-2,-1)  {};
\node [scale = .50] [circle, draw, fill = black] at (2,-1)  {};
\node [scale = .50] [circle, draw, fill = black] at (0,1)  {};
\draw (-2.25,-.5) node [scale =.85] {$p$};
\draw (2.25,-.5) node [scale =.85] {$q$};

\draw (6.1,-1) node [scale = .85] {$\tau_\text{recombination}$  };

\end{tikzpicture}
\]
\captionsetup{format=hang}
\caption{\small{The horizon problem. Without inflation the past lightcones of $p$ and $q$ never intersect. But then why does the Earth measure the same 2.7 K temperature from every direction?}}\label{horizon problem figure}
\end{figure}

\medskip
\medskip

\begin{figure}[h]
\[
\begin{tikzpicture}[scale = 0.75]

\shade(-4.5,2) -- (-4.5,-2) -- (4.5,-2) -- (4.5,2);

\draw [<->,thick] (-4.5,-2) -- (4.5,-2);

\draw (5.5, -2) node [scale = .85] {$\tau = 0$};

\draw [thick, red] (-4.5,-1) -- (4.5,-1);

\draw [dashed, thick, blue] (-2,-1) .. controls (0,-1.5) .. (4.5,-2);
\draw [dashed, thick, blue] (-2,-1) .. controls (-3,-1.25) .. (-4.5,-1.5);
\draw [dashed, thick, blue] (2,-1) .. controls (3,-1.25) .. (4.5,-1.5);
\draw [dashed, thick, blue] (2,-1) .. controls (0,-1.5) .. (-4.5,-2);

\node [scale = .50] [circle, draw, fill = black] at (-2,-1)  {};
\node [scale = .50] [circle, draw, fill = black] at (2,-1)  {};

\draw (-2.25,-.5) node [scale =.85] {$p$};
\draw (2.25,-.5) node [scale =.85] {$q$};

\draw (6.1,-1) node [scale = .85] {$\tau_\text{recombination}$  };

\end{tikzpicture}
\]
\captionsetup{format=hang}
\caption{\small{Inflation solves the horizon problem by widening the  past lightcones.}}\label{inflation solves horizon figure}
\end{figure}
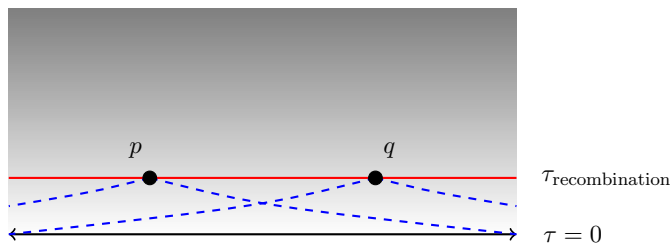

By using conformal time $\tilde{\tau}$ given by $d\tilde{\tau} = d\tau/a(\tau)$, it is an elementary exercise to show that there is no horizon problem provided the \emph{particle horizon} at the moment of last scattering is infinite: 
\begin{equation}
 \int_0^{\tau_{\text{recombination}}} \frac{1}{a(\tau)}d\tau \, = \, \infty.
\end{equation}
This condition widens the past lightcones of $p$ and $q$ so that they intersect before $\tau = 0$. See Figure \ref{inflation solves horizon figure}.

\medskip
\medskip

%

\medskip
\medskip

\begin{prop}
The particle horizon for an inflationary spacetime is infinite.
\end{prop}

\proof
From the definition of an inflationary spacetime, we have
\begin{equation}
\lim_{\tau \to 0} \frac{a(\tau)}{\tau} \, = \, 1.
\end{equation}
Therefore for any $\e > 0$ there exists a $\delta > 0$ such that $|a(\tau)/\tau - 1| < \e$ for all $0 < \tau < \delta$. Hence $1/a(\tau) > 1/(1+\e)\tau$ for all $0 < \tau < \delta$. Then the particle horizon at the moment of last scattering is 
\begin{equation}
\int_0^{\tau_\text{recombination}} \frac{1}{a(\tau)}d\tau \, \geq \,
\int_0^\delta \frac{1}{a(\tau)}d\tau \, \geq \, \int_0^\delta \frac{1}{(1+ \e )\tau}d\tau \, = \, \infty
\end{equation}
  Thus the particle horizon is infinite.
\qed

\medskip
\medskip

\begin{figure}[h]
\[
\begin{tikzpicture}[scale = 0.75]

\shadedraw [white] (-4.1,2.1) -- (0,-2) -- (4.1,2.1);
\draw [dashed, thick, blue] (0,-2) -- (4.1,2.1);
\draw [dashed, thick, blue] (0,-2) -- (-4.1,2.1);

\draw [<->,thick] (0,-3.5) -- (0,2.35);
\draw [<->,thick] (-4.5,-2) -- (4.5,-2);

\draw (-.35,2.5) node [scale = .85] {$t$};
\draw (4.75, -2.25) node [scale = .85] {$x^i$};
\draw (-.25,-2.25) node [scale = .85] {$\ms{O}$};

\draw [thick, red] (-3.5,2.1) .. controls (0, -1.3).. (3.5,2.1);

\draw [dashed, thick, blue](-1.4,-.55) -- (-1,-0.15) -- (.3,-1.45);
\draw [dashed, thick, blue](-0.3,-1.45) -- (1,-0.15) -- (1.4,-.55);

\node [scale = .50] [circle, draw, fill = black] at (-1,-.15)  {};
\node [scale = .50] [circle, draw, fill = black] at (1,-.15)  {};

\draw (-.75,0.3) node [scale =.85] {$p$};
\draw (.75,0.3) node [scale =.85] {$q$};

\draw [->] [thick]  (3.2,-0.4) arc [start angle=-90, end angle=-155, radius=40pt];
\draw (4.6,-0.4) node [scale = .85] {$\tau_\text{recombination}$  };

\end{tikzpicture}
\]
\captionsetup{format=hang}
\caption{\small{A Milne-like spacetime modeling our universe. The points $p$ and $q$ have past lightcones which intersect at some point above $\ms{O}$.}}\label{horizon figure}
\end{figure}
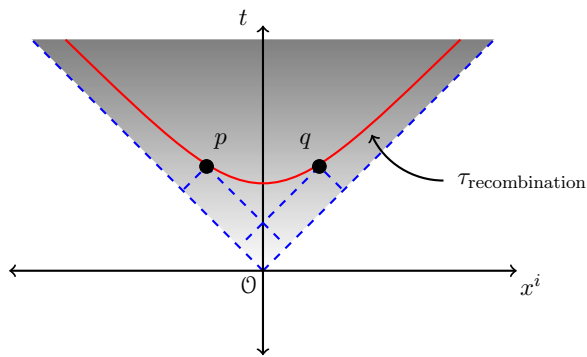

For Milne-like spacetimes, the origin $\ms{O}$ plays a unique role. The lightcones of any two points must intersect above $\ms{O}$. 
This follows from the metric being conformal to Minkowski space, $g_{\mu\nu} = \Omega^2(\tau)\eta_{\mu\nu}$. As such the lightcones are given by 45 degree angles; see Figure \ref{horizon figure} which clarifies the situation depicted in Figure \ref{inflation solves horizon figure}.

\medskip

\medskip
\medskip

Also we observe that the comoving observers all emanate from the origin $\ms{O}$. Indeed a comoving observer $\g(\tau)$ is specified by a point $(R_0, \theta_0, \phi_0)$ on the hyperboloid. 
\begin{equation}
\g(\tau) \, = \, (\tau, R_0, \theta_0, \phi_0).
\end{equation}
In the $(t, r, \theta, \phi)$ coordinates introduced in equation (\ref{t and r}), the comoving observer is given by 
\begin{equation}
\g(\tau) \, = \, \big(t(\tau), r(\tau), \theta_0, \phi_0)
\end{equation}
where 
\begin{equation}
t(\tau) \, = \, b(\tau) \cosh(R_0) \:\:\:\: \text{ and } \:\:\:\: r(\tau) \, = \, b(\tau) \sinh(R_0).
\end{equation}
Thus the relationship between $t$ and $r$ for $\g$ is $t = \coth(R_0) r$. Therefore for any comoving observer, we have $t = Cr$ for some $C > 1$. Thus the comoving observers emanate from the origin.

\begin{figure}[h]
\[
\begin{tikzpicture}[scale = 0.75]

\shadedraw [white] (-4.1,2.1) -- (0,-2) -- (4.1,2.1);
\draw [dashed, thick, blue] (0,-2) -- (4.1,2.1);
\draw [dashed, thick, blue] (0,-2) -- (-4.1,2.1);

\draw [<-,thick] (0,-3.5) -- (0,2.0);
\draw [<->,thick] (-4.5,-2) -- (4.5,-2);

\draw (4.75, -2.25) node [scale = .85] {$x^i$};
\draw (-.25,-2.25) node [scale = .85] {$\ms{O}$};
%

	\draw [thick, purple] (0,-2) -- (2,2.1);
	\draw [thick, purple] (0,-2) -- (3,2.1);
	\draw [thick, purple] (0,-2) -- (1,2.1);
	\draw [thick, purple] (0,-2) -- (-1,2.1);
	\draw [thick, purple] (0,-2) -- (-2,2.1);
	\draw [thick, purple] (0,-2) -- (-3,2.1);
	\draw [thick, purple] (0,-2) -- (0,2.1);

\end{tikzpicture}
\]
\captionsetup{format=hang}
\caption{\small{The comoving observers in a Milne-like spacetime. They all emanate from the origin $\ms{O}$.}}\label{comoving figure}
\end{figure}
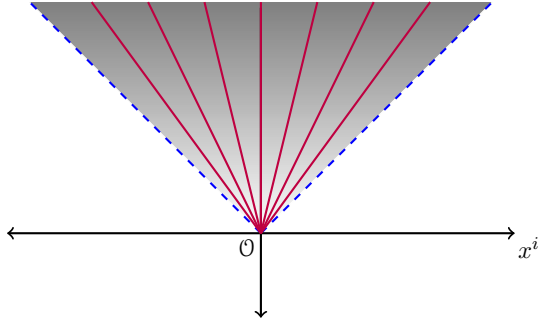

\subsection{The cosmological constant appears as an initial condition}\label{cosmo constant section}

In this section we show how the cosmological constant $\Lambda$ can appear as an initial condition for Milne-like spacetimes. This may help explain the origin of $\Lambda$. If dark energy is really modeled by a cosmological constant and not by some other model (e.g. quintessence), then $\Lambda$ would have been fixed at the big bang.

 Another interesting result in this direction is \cite{CosmoConstTop}. In their paper the authors show how the cosmological constant may arise from a topological quantity via Chern-Simons invariants. The authors make use of exotic $4$-manifolds admitting hyperbolic geometry which are ultimately expressed as hyperbolic FLRW spacetimes.

Fix $k\geq 2$. For this section let $(M,g)$ denote a $C^k$ Milne-like spacetime. Consider the Einstein equations with a cosmological constant
\begin{equation}
G_{\mu\nu} + \Lambda g_{\mu\nu}\,=\, R_{\mu\nu} - \frac{1}{2}R g_{\mu\nu} + \Lambda g_{\mu\nu} \,=\, 8\pi T_{\mu\nu}.
\end{equation}
Let $u = \pd / \pd \tau$ denote the four-velocity of the comoving observers and let $e$ be any unit spacelike orthogonal vector (its choice does not matter by isotropy). We define the \emph{normal} energy density $\rho_{\rm normal}(\tau)$ and \emph{normal} pressure function $p_{\rm normal}(\tau)$ in terms of the energy-momentum tensor
\begin{align}
\rho_{\rm normal} \,&=\, T_{\mu\nu}u^\mu u^\nu
\\
p_{\rm normal} \,&=\, T_{\mu\nu}e^\mu e^\nu
\end{align}
Then the energy density $\rho$ and pressure function $p$ in terms of $\rho_{\rm normal}$ and $p_{\rm normal}$ are given by
\begin{align}
\rho  \, &= \, \frac{1}{8\pi}G_{\mu\nu}u^\mu u^\nu 
\,=\, \rho_{\rm normal} + \frac{\Lambda}{8\pi} \label{rho eq}
\\
p \, &= \, \frac{1}{8\pi}G_{\mu\nu}e^\mu e^\nu  \,=\, p_{\rm normal} - \frac{\Lambda}{8\pi}
\end{align}
If $\rho_{\rm normal} = p_{\rm normal} = 0$ (e.g. de Sitter), then the equation of state for the cosmological constant is fixed for all $\tau$.
\begin{equation}
\rho \, = \, - p \, =\,  \frac{\Lambda}{8\pi}.
\end{equation}
We show that this equation of state appears as an initial condition. For the following theorem, we define $\rho(0) := \lim_{\tau \to 0} \rho(\tau)$. Likewise with $p(0)$ and $\rho_{\rm normal}(0)$ and $p_{\rm normal}(0)$.
\medskip
\medskip

\begin{thm}\label{initial condition thm}
Suppose the scale factor satisfies $a''(\tau) = \a \tau + o(\tau)$. Then
\[
\rho(0) \, = \, -p(0) \, = \, \frac{3}{8\pi} \a .
\]
\end{thm}

\medskip
\medskip

We prove Theorem \ref{initial condition thm} at the end of this section. First we understand its implications. If the cosmological constant $\Lambda$ is the dominant energy source during the Planck era, then we have the following connection between $\Lambda$ and the initial condition of the scale factor.

\medskip
\medskip

\begin{prop}\label{cosm const prop}
Suppose the scale factor satisfies $a''(\tau) = \a\tau + o(\tau)$,  and we have
 $\rho_{\rm normal}(0) \,=\, p_{\rm normal}(0) \,=\, 0$. Then
\[
\Lambda \, = \, 3 \a \,=\, 3a'''(0).
\]
\end{prop}

\proof
This  follows from Theorem \ref{initial condition thm} and equation (\ref{rho eq}). 
\qed

\medskip
\medskip

\noindent\emph{Remark.} In (3+1)-dimensional de Sitter space we have $T_{\mu\nu} = 0$ and $\Lambda = 3$.
In the open slicing coordinates of de Sitter, we have $a(\tau) = \sinh(\tau)$. Hence $\a = a'''(0) = 1$. Therefore de Sitter space is a special example of Proposition \ref{cosm const prop}.

\medskip
\medskip

Now we examine how an inflaton scalar field behaves in the limit $\tau \to 0$. We will demonstrate that slow-roll inflation follows if the initial condition for the potential is given by the cosmological constant:  $V|_{\tau = 0} = \Lambda/8\pi$.
 Recall the energy-momentum tensor for a scalar field $\phi$ is
\begin{equation}
T^\phi_{\mu\nu} \, = \, \nabla_\mu \phi \nabla_\nu \phi \,-\, \left[\frac{1}{2}\nabla^\s \phi \nabla_\s \phi \,+\, V(\phi) \right]g_{\mu\nu}.
\end{equation}
And its energy density and pressure function are
\begin{equation}
\rho_\phi (\tau) \,=\, \frac{1}{2}\phi '(\tau)^2 \,+\, V\big(\phi(\tau)\big) \:\:\:\:\:\: \text{ and } \:\:\:\:\:\: p_\phi (\tau) \,=\, \frac{1}{2}\phi '(\tau)^2 \,-\, V\big(\phi(\tau)\big).
\end{equation}

\medskip

The following proposition shows that when the initial condition for $V$ is determined by the cosmological constant, then one obtains a slow-roll era.

\medskip
\medskip

\begin{prop}
Suppose the scale factor satisfies $a''(\tau) = \a\tau + o(\tau)$, and we have 
 \[\rho \,\to\, \rho_\phi \:\:\:\:\:\: \text{ and } \:\:\:\:\:\: p\,\to\, p_\phi \:\:\:\:\:\: \text{ as } \: \tau \,\to\, 0. \]
\begin{itemize}

\item[\emph{(1)}] If $V\big(\phi(0)\big) = 3\a/8\pi$, then $\phi'(0) = 0$. 

\item[\emph{(2)}] If $\rho_{\emph{\text{normal}}}(0) = p_{\emph{\text{normal}}}(0) =0$ and $V\big(\phi(0)\big) = \Lambda/8\pi$, then $\phi'(0) = 0$

\end{itemize}
\end{prop}

\proof
(1) follows from Theorem \ref{initial condition thm} and (2) follows from  Proposition \ref{cosm const prop}.
\qed

\medskip
\medskip
\medskip
\medskip

\noindent\underline{\emph{Proof of Theorem \emph{\ref{initial condition thm}}}}:

\medskip

Friedmann's equations are $(8\pi/3)\rho = H^2 -1/a^2$ and $8\pi p = -2a''/a - (8\pi/3)\rho$ where $H = a'/a$ is the Hubble parameter. Using  $a(\tau) = \tau + f(\tau)$, the Friedmann equations become
\begin{equation}
\frac{8\pi}{3} \rho(\tau) = \left(\frac{a'(\tau)}{a(\tau)}\right)^2 - \frac{1}{a(\tau)^2} = \frac{2f'(\tau) + f'(\tau)^2}{\big[\tau + f(\tau)\big]^2} = \frac{\big(f'(\tau)/\tau\big)\big[2/\tau + f'(\tau)/\tau\big]}{\big(1 + f(\tau)/\tau\big)^2}
\end{equation}
and
\begin{equation}
-8\pi p(\tau) = 2\frac{a''(\tau)}{a(\tau)} + \frac{8\pi}{3}\rho(\tau) = \frac{2f''(\tau)/\tau}{1 + f(\tau)/\tau} + \frac{8\pi}{3}\rho(\tau).
\end{equation}

By definition of an inflationary spacetime, we have $f'(0) := \lim_{\tau \to 0}f(\tau)/\tau = 0$. Also, since $a''(\tau) = \a\tau + o(\tau)$, we have $0 = a''(0) = f''(0) = \lim_{\tau \to 0}f'(\tau)/\tau$ and $\a = \lim_{\tau \to 0}f''(\tau) /\tau$. Therefore for all $\e > 0$, there is a $\delta > 0$ such that $|f''(\tau)/\tau - \a| < \e$ for all $0 < \tau < \delta$. Integrating this expression gives $(\a - \e)\tau/2 < f'(\tau)/\tau < (\a + \e)\tau/2$. Plugging this into the first Friedmann equation yields $8\pi\rho(0)/ 3 = \a$. Using this for the second Friedmann equation yields $-8\pi p(0) = 3\a$.
\qed

\subsection{Lorentz invariance}\label{Lorentz invariance section}

\medskip

In this section we show that the isometry group for Milne-like spacetimes contains the Lorentz group. Since Lorentz invariance plays a pivotal role in QFT (e.g. the field operators are constructed out of finite dimensional  irreducible represenations of the Lorentz group \cite{Weinberg, Tung}), Milne-like spacetimes are a good background model if one wants to develop a quantum theory of cosmology.

\medskip
\medskip

\noindent{\emph{Remark.}} In this section $\Lambda$ will always denote an element of the Lorentz group (i.e. a Lorentz transformation) and not the cosmological constant. 

\medskip
\medskip

Let $\eta_{\mu\nu}$ be the Minkowski metric. The \emph{Lorentz group} is
\begin{equation}
\text{L} \, = \, \rm{O}(1,3) \,= \, \{\Lambda \mid \eta_{\mu\nu} = \Lambda^\a_{\:\:\:\mu}\Lambda^\b_{\:\:\:\nu}\eta_{\a\b} \}.
\end{equation}
A Lorentz transformation $\Lambda$ shifts elements in Minkowski space via $x^\mu \mapsto \Lambda^\mu_{\:\:\:\nu}x^\nu$, but it leaves the hyperboloids fixed. More generally this applies to any Milne-like spacetime by the same map.

\begin{figure}[h]
\[
\begin{tikzpicture}[scale = 0.75]

\shadedraw [white] (-4.1,2.1) -- (0,-2) -- (4.1,2.1);
\draw [dashed, thick, blue] (0,-2) -- (4.1,2.1);
\draw [dashed, thick, blue] (0,-2) -- (-4.1,2.1);

\draw [<->,thick] (0,-3.5) -- (0,2.35);
\draw [<->,thick] (-4.5,-2) -- (4.5,-2);

\draw (-.35,2.5) node [scale = .85] {$t$};
\draw (4.75, -2.25) node [scale = .85] {$x^i$};
\draw (-.25,-2.25) node [scale = .85] {$\ms{O}$};

\draw [thick, red] (-3.5,2.1) .. controls (0, -1.3).. (3.5,2.1);

\node [scale = .60] [circle, draw, fill = black] at (-2.5,1.1)  {};
\node [scale = .60] [circle, draw, fill = black] at (.8, -.26)  {};
\draw (-3,0.3) node [scale =.85] {$p$};
\draw (.5,-1) node [scale =.85] {$q$};

\end{tikzpicture}
\]
\captionsetup{format=hang}
\caption{\small{A Lorentz transformation $\Lambda$ based at $\ms{O}$ shifts points $p$ to other points $q = \Lambda p$ on the same $\tau = $ constant slice. For Milne-like spacetimes, $\Omega$ is a function of $\tau$. Therefore $\Omega (\tau) = \Omega(\tau \circ \Lambda)$, e.g. in this figure we would have $\Omega \big(\tau (p)\big) = \Omega\big(\tau(q)\big)$.}}\label{Lorentz figure}
\end{figure}
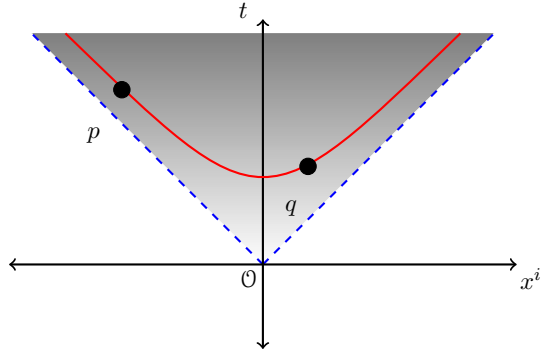

For a Milne-like spacetime, we have $g_{\mu\nu} = \Omega^2(\tau) \eta_{\mu\nu}$ where $\eta_{\mu\nu}$ is the usual Minkowski metric. Since a Lorentz transformation leaves hyperboloids invariant, we have 
\begin{equation}
\Omega(\tau) \, = \, \Omega(\tau \circ \Lambda).
\end{equation}

Recall the Lorentz group $\text{L} = \text{O}(1,3)$ has four connected components $\text{L}^\uparrow_{+}$, $\text{L}^\uparrow_-$, $\text{L}^\downarrow_{+}$, $\text{L}^\downarrow_{-}$. The $\pm$ corresponds to $\det \Lambda = \pm 1$, the $\uparrow$ corresponds to $\Lambda^0_{\:\:\:0} \geq 1$, and the $\downarrow$ corresponds to $\Lambda^0_{\:\:\:0} \leq -1$. 

Lorentz transformations fix the origin (i.e. $\Lambda \ms{O} = \ms{O}$) and are isometries on the spacetime manifold with boundary $(M \cup \pd^-{M}) \setminus \{\ms{O}\}$. We will say that any map which fixes $\ms{O}$ and is an isometry on the spacetime manifold with boundary $(M \cup \pd^-M) \setminus\{\ms{O}\}$ is an \emph{isometry on $M \cup \pd^-M$.} Note that the set of isometries on $M \cup \pd^-M$ forms a group via composition. Since Milne-like spacetimes are defined for $t > 0$, only the subgroup $\text{L}^\uparrow = \text{L}^\uparrow_+ \cup \text{L}^\uparrow_-$ acts by isometries on Milne-like spacetimes. If $(M,g)$ admits a $C^2$-extension, then we obtain an isomorphism.

\medskip
\medskip

\begin{thm}\label{isometry thm}
Let $(M,g)$ be a Milne-like spacetime. Then any $\Lambda \in \emph{\text{L}}^\uparrow$ is an isometry on $M \cup \pd^-M$.
\end{thm}

%
%

\medskip
\medskip

\begin{thm}\label{milne isometry}
If a Milne-like spacetime admits a $C^2$-extension, then  $\emph{\text{L}}^\uparrow$ is isomorphic to the group of isometries on $M \cup \pd^-M$.
\end{thm}

\medskip
\medskip

\noindent\emph{Remark.} To the best of the author's knowledge, Theorem \ref{milne isometry} is a new result. Its proof relies on the existence of $\pd^-M$.

\medskip
\medskip

\noindent\underline{\emph{Proofs of Theorems \emph{\ref{isometry thm}} and \emph{\ref{milne isometry}}}}:

\medskip

Let $\Lambda$ be an element of $\text{L}^\uparrow$. It produces a unique map, $x \mapsto \Lambda x$ via $x^\mu \mapsto \Lambda^\mu_{\:\:\:\nu}x^\nu$ where $(x^0, x^1, x^2, x^3) = (t, x, y, z)$ are the conformal Minkowski coordinates introduced in the proof of Theorem \ref{extension thm}. Since $(M,g)$ is only defined for $t > 0$, we must restrict to Lorentz transformations $\Lambda \in \text{L}^\uparrow$. Consider a point $p \in M$ and a tangent vector $X = X^\mu \pd_\mu$ at $p$. Then $\Lambda$ acts on $X$ by $d\Lambda (X) = \Lambda^\mu_{\:\:\:\nu} X^\nu \pd_\mu$ and sending it to the point $\Lambda p \in M$. Since $\Omega(\tau \circ p) = \Omega(\tau \circ \Lambda p)$, we have
\begin{align*}
g_{\mu\nu}(d\Lambda X)^\mu (d \Lambda Y)^\nu \, &= \, \Omega^2(\tau \circ\Lambda p) \, \eta_{\mu\nu} (d\Lambda X)^\mu (d \Lambda Y)^\nu
\\
&= \, \Omega^2(\tau \circ p) \, \eta_{\mu\nu}(\Lambda^\mu_{\:\:\:\a}X^\a)(\Lambda^\nu_{\:\:\: \b}Y^\b)
\\
&= \, \Omega^2( \tau \circ p)\, \eta_{\a \b} X^\a Y^\b
\\
&= \, g_{\a\b}X^\a Y^\b.
\end{align*}
 Thus $\Lambda$ is an isometry. Now consider $p \in \pd^-M$. Then $\Lambda p \in \pd^-M$ and we have $\Omega|_p = \Omega|_{\Lambda p} = \Omega(0)$. Therefore the calculation above carries through in this case as well. This proves Theorem \ref{isometry thm}.

 Now we prove Theorem \ref{milne isometry}. By Theorem \ref{isometry thm} we have $\text{L}^\uparrow$ is a subgroup, so it suffices to show it's the whole group.
 Suppose $f$ is an  isometry  on $M \cup \pd^-M$. The differential map $df_{\ms{O}}$ is a linear isometry on the tangent space at $\ms{O}$. Therefore $df_{\ms{O}}$ corresponds to an element of the Lorentz group, say $\Lambda^\mu_{\:\:\:\nu}$. It operates on vectors $X$ at $\ms{O}$ via $df(X) = \Lambda^\mu_{\:\:\:\nu}X^\nu\pd_\mu$. Now we define the isometry $\tilde{f}$ by $\tilde{f}(x) = \Lambda^\mu_{\:\:\:\nu}x^\nu$. Consider the set 
\[
A \,=\, \{p \in M \cup \pd^-M \,\mid \, df_p = d\tilde{f}_p\}.
\]
Note that if $df_p = d\tilde{f}_p$, then $f(p) = \tilde{f}(p)$. Hence it suffices to show $A = M \cup \pd^-M$. $A$ is nonempty since $\ms{O} \in A$, and $A$ is closed because $df - d\tilde{f}$ is continuous. So since $M \cup \pd^-M$ is connected, it suffices to show $A$ is open in the subspace topology. 
Let $p \in A$. Since $\Omega$ is $C^2$, there is a normal neighborhood $U$ about $p$. If $q \in U$, there is a vector $X$ at $p$ such that $\exp_p(X) = q$. Since isometries map geodesics to geodesics, they satisfy the property $f \circ \exp_p = \exp_{f(p)} \circ \, df_p$ for all points in $U$ (see page 91 of \cite{ON}). Therefore
\[
f(q) = f\big(\exp_{p}(X)\big) = \exp_{f(p)}(df_pX) = \exp_{\tilde{f}(p)}(d\tilde{f}_pX) = \tilde{f}\big(\exp_p(X)\big) = \tilde{f}(q).
\]
Thus $f(q) = \tilde{f}(q)$ for all $q \in U$; hence $df_q = d\tilde{f}_q$ for all $q \in U$. Therefore $A$ is open.
\qed

\medskip
\medskip

\subsection{A possible dark matter particle?}

\medskip

The symmetries in quantum theory can be characterized by local symmetries  and global spacetime symmetries. 

Local symmetries correspond to the gauge symmetry group $\text{SU}(3) \times \text{SU}(2) \times \text{U}(1)$ of the standard model. In gauge theory the Lagrangian is invariant under position-dependent gauge transformations; hence the world `local.' Local gauge invariance necessitates the existence of gauge fields in the Lagrangian which are then checked experimentally. The $\text{SU}(3)$ part describes the strong interaction of quantum chromodynamics. The $\text{SU}(2) \times \text{U}(1)$ part describes the electroweak interaction.

Global spacetime symmetries are the symmetries of the underlying spacetime manifold. Since the standard model is modeled on Minkowski space, the global spacetime symmetry group is the Poincar{\'e} group which are the isometries in Minkowski space. Wigner's classification \cite{Wigner_classification} of the irreducible unitary representations of the Poincar{\'e}  group described the spin properties of elementary particles which is considered a huge success in mathematical quantum field theory.

If one wants to build a quantum theory on a cosmological background, then Milne-like spacetimes are a preferred model since they are Lorentz invariant. Wigner's success in the classification of the Poincar{\'e} group motivates us to seek the irreducible unitary representations of the Lorentz group. Similar to Wigner's analysis, we desire all projective unitary representations to lift to unitary representations. Therefore we really seek the irreducible unitary representations of $\text{SL}(2,\C)$ which is the simply connected double cover of $\text{L}^\uparrow_+$.

\medskip
\medskip

\noindent{\bf Classification of the irreducible unitary representations of $\text{SL}(2,\C)$:} 

\noindent This classification comes from Theorem 10.9 in \cite{Tung}. There are two classes of irreducible unitary representations of $\text{SL}(2,\C)$. The first class is the \emph{principal series}. These particles are characterized by the parameter $\nu = -iw$ where $w$ is real and spin $j = 0, 1/2, 1, \dotsc$ The second class is the \emph{complementary series}. These particles are characterized by the parameter $-1 \leq \nu \leq 1$ and spin $j = 0$. The terminology `principal' and `complementary' comes from the classification of irreducible unitary representations of semi-simple Lie groups \cite{Knapp}. 

\medskip
\medskip

Given that the comoving observers in a Milne-like spacetime all emanate from the origin $\ms{O}$ (see Figure \ref{comoving figure}), a possible physical interpretation of this classification would be that these are the particles created at the big bang.  Then the principal series would correspond to the particles which make up the standard model (in analogy to Wigner's classification of the Poincar{\'e} group), but this leaves the complementary series up to interpretation. Perhaps
\begin{align*}
\text{complementary series} \,&=\, \text{dark matter particles?} 
\end{align*}

\medskip

But is there any evidence that dark matter is comprised of spin 0 particles? Yes. Scalar field dark matter (SFDM)  \cite{GuzmanMatos, GuzmanLopez,  MaganaMatos, AbrilRoblesMatos} also known as Bose-Einstein condensate (BEC) dark matter \cite{SeidelSuen, Sin, SinJi, MatosLopez, LopezArgelia, SKM} also known as wave dark matter (WDM) \cite{Bray1, Bray2, GoetzThesis, ParryThesis} also known as fuzzy dark matter (FDM) \cite{FDM1, FDM2} all use the Klein-Gordon equation (i.e. the wave equation for spin 0 particles) to model dark matter. The difference in name comes from a difference in motivation. One reason for introducing models of dark matter based on the Klein-Gordon equation is to alleviate the cusp problem associated with the weakly interacting massive particle (WIMP) models of dark matter \cite{LeeDarkMatter}. Furthermore, the models based on the Klein-Gordon equation reproduce the observed spiral pattern density in disk galaxies (see Figures 1 - 4 in \cite{Bray1}) which makes these models promising.

The parameters $\nu$ and $j$ in the classification of $\text{SL}(2,\C)$ are determined by Casimir operators built out of rotations and Lorentz boosts (see equation 10.3-1 in \cite{Tung}). The spin parameter $j$ comes from the usual rotation generators $[J_i, J_j] = i\e^{ijk}J_k$. The parameter $\nu$ is determined by requiring unitary of the boost generators $[K_i, K_j] = -i\e^{ijk}J_k$. See section 10.3.3 and appendix VII of \cite{Tung} for the full details.

What's interesting is that the parameter $\nu$ takes on very different forms for the principal series and the complementary series. 
It would be interesting if there is any new physics here. If the identification ``principal series = normal matter" and ``complementary series = dark matter" is true, then the distinguishing feature could be related to this parameter $\nu$. Perhaps this could offer an explanation for dark matter's lack of interaction with electromagnetism.

\medskip
\medskip

\subsection{What lies beyond $\tau = 0$?}\label{antimatter section}

\medskip

Since Milne-like spacetimes extend through $\tau = 0$, it is an interesting question to ask what exists in the extension. Of course this is only speculation, but hints can be found when one considers the maximal analytic extension whenever $\Omega$ is analytic on $M \cup \pd^-M$.  For $a(\tau) = \tau$ (i.e. the Milne universe), the maximal analytic extension is Minkowksi space. For $a(\tau) = \sinh(\tau)$, the maximal analytic extension is de Sitter space.

For Minkowski space and de Sitter space, we have the full Lorentz group $\text{L} = \text{O}(1,3)$ acting as isometries at the origin $\ms{O}$. When elements in $\text{L}^\downarrow = \text{L}^\downarrow_+ \cup \text{L}^\downarrow_-$ act at the origin, it produces a PT symmetric spacetime (i.e. one where the map $(t,x,y,z) \mapsto (-t,-x,-y,-z)$ is an isometry).

\medskip
\medskip

\noindent{\bf Lorentz invariance at $\ms{O}$ implies an antimatter universe}

\medskip

Let $(M,g)$ be a Milne-like spacetime. Requiring $\ms{O}$ to be Lorentz invariant (i.e. the full Lorentz group $\text{L} = \text{O}(1,3)$ acts at the origin $\ms{O}$) produces a PT symmetric universe. Given the CPT theorem \cite{PCT}, perhaps the universe's missing antimatter is contained in the PT symmetric universe.

We remark this idea is closely related to the same idea in \cite{TBF}. There the authors consider a $k = 0$ FLRW spacetime with metric $g = -d\tau^2 + a^2(\tau)\big[dx^2 + dy^2 + dz^2\big]$ in a radiation dominated era $a(\tau) \propto \sqrt{\tau}$. By moving to conformal time $\tilde{\tau}$ given by $d\tilde{\tau} = d\tau/a(\tau)$, one arrives at the metric $g = a^2(\tilde{\tau})\big[-d\tilde{\tau}^2 + dx^2 + dy^2 + dz^2 \big]$ where $a(\tilde{\tau}) \propto \tilde{\tau}$. They then analytically extend the function $a(\tilde{\tau})$ from $(0, + \infty)$ to $\R$ and call the $(-\infty, 0)$ part the `CPT-symmetric' universe. However, at $\tilde{\tau} = 0$, the metric is $g = 0$. Hence it's degenerate. Therefore this is not a spacetime extension.

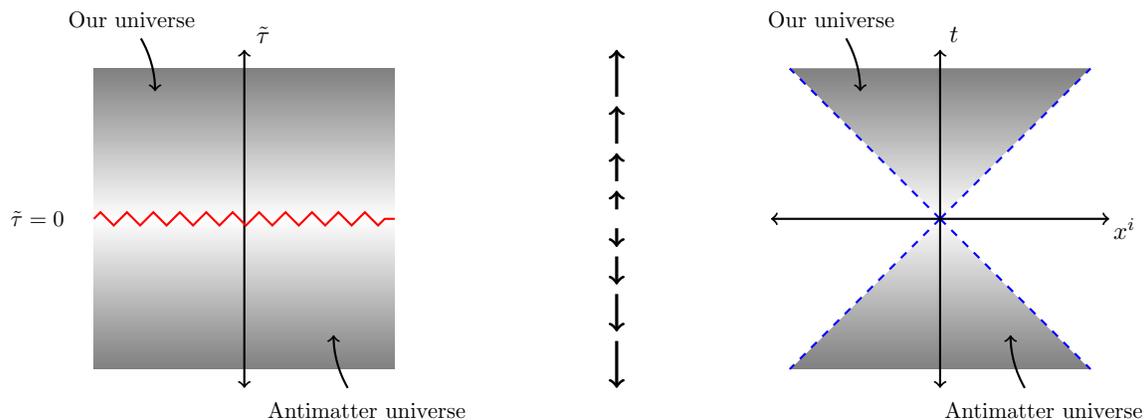
\begin{figure}[h]
\[
\begin{tikzpicture}[scale = 0.5]

\shadedraw [white]	(5.5,2) -- (9.5,-2) -- (13.5,2);
\draw [dashed, thick, blue] (9.5,-2) -- (5.5,2);
\draw [dashed, thick, blue] (9.5,-2) -- (13.5,2);
	
\shadedraw[top  color = white, bottom color = gray, white] (5.5,-6) -- (9.5,-2) -- (13.5,-6);	
\draw [dashed, thick, blue] (9.5,-2) -- (5.5,-6);
\draw [dashed, thick, blue] (9.5,-2) -- (13.5,-6);

\draw [<->,thick] (9.5,-6.5) -- (9.5,2.5);
\draw [<->,thick] (5.0,-2) -- (14.0,-2);

\draw (9.870,2.90) node [scale = .85] {$t$};
\draw (14.40, -2.25) node [scale = .85] {$x^i$};


\shade(-13,2) -- (-13,-2) -- (-5,-2) -- (-5,2);

\shade[top  color = white, bottom color = gray](-13,-2) -- (-13,-6) -- (-5,-6) -- (-5,-2);

\draw [<->,thick] (-9,-6.5) -- (-9,2.5);

\draw (-8.5,2.90) node [scale = .85] {$\tilde{\tau}$};

\draw[snake=zigzag, red, thick] (-13,-2) -- (-5,-2);

\draw (-14.5,-2) node [scale = .85] {\small{$\tilde{\tau} = 0$}};


\draw [->] [thick] (7.0,2.8) arc [start angle=30, end angle=0, radius=80pt];

\draw (6.6,3.3) node [scale = .85] {\small{Our universe}};
	
	\draw [->] [thick] (11.75,-6.5) arc [start angle=-150, end angle=-180, radius=80pt];	

\draw (12.25,-7.1) node [scale = .85] {\small{Antimatter universe} };


\draw [->] [thick] (-11.75,2.8) arc [start angle=30, end angle=0, radius=80pt];

\draw (-12,3.3) node [scale = .85] {\small{Our universe}};

\draw [->] [thick] (-6.25,-6.5) arc [start angle=-150, end angle=-180, radius=80pt];	

\draw (-5.75,-7.1) node [scale = .85] {\small{Antimatter universe}};


\draw [->, very thick] (.9,1.25) -- (.9,2.5);

\draw [->, very thick] (.9,0) -- (.9,1.0);

\draw [->, very thick] (.9,-1) -- (.9,-0.25);
	
\draw [->, very thick] (.9,-1.75) -- (.9,-1.25);	
	
	
\draw [->, very thick] (.9,-2.25) -- (.9,-2.75);

\draw [->, very thick] (.9,-3) -- (.9,-3.75);

\draw [->, very thick] (.9,-4) -- (.9,-5);
	
\draw [->, very thick] (.9,-5.25) -- (.9,-6.5);	
	
\end{tikzpicture}
\]
\caption{\small{ 
The figure on the left represents the universe/antimatter universe pair in \cite{TBF}. The metric is degenerate at $\tilde{\tau} = 0$, so the pair together do not form a spacetime. The figure on the right represents the universe/antimatter universe pair for a Milne-like spacetime. In this case the pair coexist in a single nondegenerate spacetime. The arrows in the middle represent the arrow of time determined by increasing entropy. The idea of a universe with an arrow of time opposite of ours is not new. For example see figure 9 in  \cite{CarrollJennifer}.
}}
\label{matter-antimatter}
\end{figure}

\medskip
\medskip

\noindent\emph{Remark.} The antimatter universe in Figure \ref{matter-antimatter} was speculated by assuming $\text{L}^{\downarrow}$ acts by isometries. Similarly one can speculate what lies in the white region between the universe and antimatter universe. If the Lorentz group acts by isometries, then the white region will be foliated by three-dimensional de-Sitter spacetimes.

\medskip
\medskip

\noindent{\bf How can we interpret the PT symmetric universe as an antimatter universe?}

\medskip

Consider an experimentalist. If the universe is modeled by a Milne-like spacetime, then the experimentalist will use coordinates that coincide with the comoving observers (except for a small correction due to the Milky Way's velocity relative to the CMB). Therefore the experimentalist will build rods and clocks which measure distances and times with respect to the metric 
\begin{equation}
g \,=\, \Omega^2(\tau)\big[-dt^2 + dx^2 + dy^2 + dz^2\big].
\end{equation}

\medskip
\medskip

With these coordinates, the experimentalist will measure an energy $p^0$ and momentum $p^i$ of a particle with mass $m$ such that $
 -g_{\mu\nu}p^\mu p^\nu = -\Omega^2(\tau)\eta_{\mu\nu}p^\mu p^\nu = m^2.
$
Assuming the de Broglie relations $p^\mu \to i\pd^\mu$ for a spin 0 field $\psi$ yields a Lorentz invariant Klein-Gordon equation for Milne-like spacetimes 
$
\big[\Omega^2(\tau)\eta^{\mu\nu}\pd_\mu\pd_\nu\big] \psi = m^2 \psi.
$
Likewise, for a Dirac spinor $\psi$, we have a Lorentz invariant Dirac equation for Milne-like spacetimes  

\begin{equation}\label{dirac for milne}
\big[\Omega(\tau)\gamma^\mu\pd_\mu\big] \psi \,=\, m\psi.
\end{equation}
(Recall we are using the $(-, +, +, +)$ signature convention, so our Dirac equation does not include a factor of $i$, but our $\g^\mu$ matrices do.) Lorentz invariance follows because both $\Omega$ and the original Dirac equation are Lorentz invariant.

In the Weyl representation, the matrices $\g^\mu$ are
\begin{equation}
\g^0= 
i\left(\begin{array}{cc} 0 & I\\ I & 0 \end{array}\right), \:\:\:\: \g^j = i\left(\begin{array}{cc} 0 & \s^j\\ -\s^j & 0 \end{array}\right).
\end{equation}
where $\s^j$ are the usual Pauli spin matrices.
 
 \begin{equation}
I = \s^0 =
\left(\begin{array}{cc} 1 & 0\\ 0 & 1 \end{array}\right) \:\:\:\: 
\s^1= 
\left(\begin{array}{cc} 0 & 1\\ 1 & 0 \end{array}\right)\:\:\:\: 
\s^2= 
\left(\begin{array}{cc} 0 & -i\\ i & 0 \end{array}\right)\:\:\:\: 
\s^3= 
\left(\begin{array}{cc} 1 & 0\\ 0 & -1 \end{array}\right)
\end{equation}

 Choose a single comoving observer in the PT extension. Pick coordinates $(t,x,y,z)$ so that they are aligned with this comoving observer (this can always be done by a Lorentz transformation at $\ms{O}$). Then $x = y = z =0$ along the observer's path. From equation (\ref{t and r}), the relationship between $\tau$ and $t$ is $b(\tau) = t$. Therefore $\pd_t = b'(\tau)\pd_\tau$. Hence $\pd_\tau = \Omega(\tau)\pd_t$, and so equation (\ref{dirac for milne}) becomes

\begin{equation}\label{beautiful dirac eq}
i\pd_\tau \psi \,=\, m \psi.
\end{equation}

\medskip

\noindent\emph{Remark.} There is a certain elegance in the simplicity of equation (\ref{beautiful dirac eq}). In this form we see the Dirac equation is manifestly Lorentz invariant, and it highlights a correspondence between  mass  and the proper time of comoving observers.  This suggests that the global spacetime symmetry group for Milne-like spacetimes should be $\R \times \text{O}(1,3)$ where the $\R$ factor corresponds to translations in cosmic time $\tau$ and yields the physical quantity mass. Compare this with the Poincar{\'e} group $\R^4 \rtimes \text{O}(1,3)$ which is the global spacetime symmetry group for Minkowski space. The mass for the Poincar{\'e} group comes from the $\R^4$ factor in the semi-direct product. See \cite{Milne_and_Symmetries} for a mathematical discussion of this comparison. 

\medskip
\medskip

Experimentalists in our universe $I^+(\ms{O})$ would use coordinates $(t,x,y,z)$ to make \linebreak observations and measurements. Anti-experimentalists in the anti-universe $I^-(\ms{O})$ would use coordinates $(-t,-x,-y,-z)$. Therefore the Dirac equation that the anti-experimentalists would use  is $\big[\Omega(\tau)\g^\mu (-\pd_\mu)\big]\psi = m\psi$. Note that this is equivalent to $\Omega(\tau)\g^\mu\pd_\mu \psi = -m\psi$ (i.e. it's the Dirac equation with negative mass). This explains the $\pm$ ambiguity one arrives at when deriving the Dirac equation.

 Whether $\psi$ solves the Dirac equation for $I^+(\ms{O})$ or $I^-(\ms{O})$, the anticommutation Clifford relations imply 
\begin{equation}
\Omega^2(\g^\mu\pd_\mu)(\g^\nu\pd_\nu)\psi \,=\, \Omega^2\big(\g^\mu(-\pd_\mu)\big)\big(\g^\nu(-\pd_\nu)\big) \psi \,=\, m^2\psi.
\end{equation}

We can introduce electromagnetism in the Dirac equation via an electromagnetic potential $A_\mu$ with the usual prescription $i\pd_\mu \to i\pd_\mu - eA_\mu$, or equivalently, $\pd_\mu \to \pd_\mu + ieA_\mu$. Then the corresponding Dirac equations for the experimentalist in $I^+(\ms{O})$ and the anti-experimentalist in $I^-(\ms{O})$ are, respectively,

\begin{align}\label{dirac solve with pot}
\Omega \g^\mu(\pd_\mu + i e A_\mu) \psi \,=\, m \psi
\\
\Omega \g^\mu(-\pd_\mu + i e A_\mu) \psi \,=\, m \psi
\end{align}

Define the matrices
\begin{equation}
\g(x) \,=\, \sum_{\mu =0}^3 x^\mu\g^\mu \,=\,
i\left(\begin{array}{cc} 0 & \underline{x}\\ \underline{\text{P}x} & 0 \end{array}\right) 
\end{equation}
where $\underline{x} = - x^\mu \s^\nu \eta_{\mu\nu}$ and $\text{P}x = (x^0, -x^1, -x^2, -x^3)$. Let $\text{PT} \in \text{GL}(4,\C)$ be an element which reverses both space and time. There are two choices which differ by a negative sign. We choose
\begin{equation}
\text{PT} \,=\, \left(\begin{array}{cc} I & 0 \\ 0 & -I \end{array}\right).
\end{equation}
Then $\g(-x) = \text{PT}\, \g(x)(\text{PT})^{-1}$. Hence $\text{PT}$ reverses space and time by acting on $\g(x)$ via conjugation. Note that $ \text{PT}\,\g^\mu = - \g^\mu\;\text{PT}$ and $(\g^\mu)^* = -\g^\mu$. Therefore matrix multiplication and complex conjugation yield the following table.

\medskip
\medskip
\medskip
\medskip

\begin{center}
 \begin{tabular}{c| c |c} 
 
 Spinor field & Equation & An interpretation   \\ [.75ex] 
 \hline
 & & 
 \\
 $\psi$ & $\Omega \g^\mu(\pd_\mu + ieA_\mu)\psi \,=\, m\psi$ & $\psi$ in $I^+\ms{O})$
  \\ [.75ex] 

 $\psi^*$ & $\Omega \g^\mu(-\pd_\mu + ieA_\mu)\psi^* \,=\, m\psi^*$ & $\psi$ in $I^-(\ms{O})$ 
  \\[.75ex] 

 $\text{PT}\,\psi$ & $\Omega \g^\mu(-\pd_\mu - ieA_\mu)\text{PT}\,\psi \,=\, m\text{PT}\,\psi$ & Anti $\psi$ in $I^-(\ms{O})$  
 \\ [.75ex] 
 
 $\text{PT}\,\psi^*$ & \:\: $\Omega \g^\mu(\pd_\mu - ieA_\mu)\text{PT}\,\psi^* \,=\, m\text{PT}\,\psi^*$ \:\: & Anti $\psi$ in $I^+(\ms{O})$ 
 \\ [.75ex] 

\end{tabular}
\end{center}

\medskip
\medskip
\medskip
\medskip

Given the interpretation, perhaps the big bang at $\ms{O}$ produced equal amounts of matter and antimatter. The matter, represented by $\psi$, traveled into our universe $I^+(\ms{O})$ while the antimatter, represented by $\text{PT}\,\psi$, traveled into the antimatter universe $I^-(\ms{O})$. The antimatter that we observe in our universe comes in the form $\text{PT}\,\psi^*$ while the anti-antimatter the anti-experimentalists observe comes in the form $\psi^*$.

\medskip
\medskip

\noindent\emph{Remark.} The Lorentz invariance of the Dirac equation described in this section can be carried over to  Lagrangians in QFT by appropriately including factors of $\Omega.$

\medskip
\medskip

\appendix

\section{A Brief Account of Cosmological Singularity  Theorems}\label{singularity theorem appendix}

\medskip

In this appendix we give a brief account of some of the singularity theorems used in cosmology. The purpose of this appendix is to demonstrate that the singularity theorems don't always apply to inflationary spacetimes. Indeed Milne-like spacetimes can almost always be used as counterexamples.

The first step in developing a cosmological theory  is to assume the Copernican principle. This assumption is supported by the highly uniform CMB radiation. The Copernican principle implies that the spacetimes $(M,g)$ which model cosmology are  given by $M = I \times \S$ and $g = -d\tau^2 + a^2(\tau)h$ where $I\subset \R$ is an interval and $(\S,h)$ are spaces of constant sectional curvature (i.e. maximally symmetric spaces). These are called FLRW spacetimes. Let $\tau_{0} = \inf I$. If one assumes the universe is in a radiation-dominated era for all $\tau_{0} < \tau < \tau_1$ given some $\tau_1$, then one finds $a(\tau) \to 0$ as $\tau \to \tau_{0}$ and $\tau_{0} > -\infty$. In this case we say $\tau_{0}$ is the big bang. By shifting the $\tau$ coordinate, we can assume $\tau_0 = 0$. Moreover the scalar curvature diverges as $\tau \to 0$, so $\tau = 0$ admits a curvature singularity.
These arguments generalize if one replaces the assumption that the universe is in a radiation-dominated era with the assumption that the universe is nonvacuum and obeys the strong energy condition. 

The singularity theorems of Hawking and Penrose \cite{HE} demonstrated  that singularities (in the sense of timelike or null geodesic incompleteness) are a generic feature of physically relevant spacetimes. These theorems don't assume any symmetry conditions on the spacetime manifold, but they do assume the strong energy condition.  Hawking's cosmological singularity theorems (see Theorems 55A and 55B in \cite{ON}) both assume the strong energy condition. 

There is a problem with the strong energy condition assumption in the singularity theorems. Assuming this condition in our universe, one finds that the particle horizon is finite. This implies that there are parts of the CMB that never achieved causal contact in the past. But if this is true, then how could the CMB have such a perfectly uniform temperature? This became known as the horizon problem in cosmology \cite{WeinbergCos}. 

A resolution to the horizon problem is to assume that the universe underwent a brief period of accelerated expansion, $a''(\tau) > 0$, immediately after the big bang and right before the radiation-dominated era. This would allow for causal contact between the different points on the CMB. This theory became known as inflationary theory and was first put forth by Alan Guth \cite{GuthInflation}. It also solved the flatness problem of cosmology and the magnetic monopole problem of certain grand unified theories \cite{WeinbergCos}.

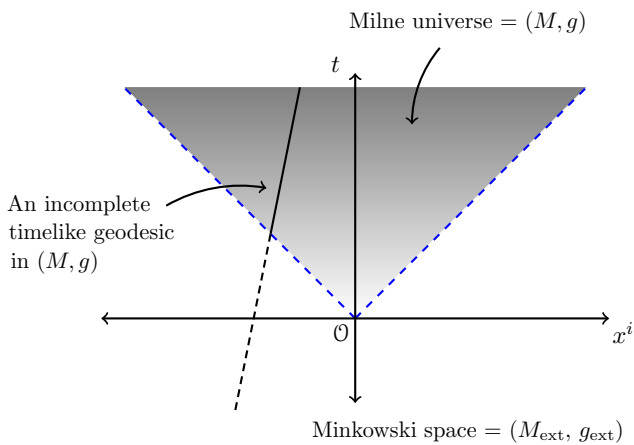
\begin{figure}[h]
\[
\begin{tikzpicture}[scale = .75]

\shadedraw [white] (-4.1,2.1) -- (0,-2) -- (4.1,2.1);
\draw [dashed, thick, blue] (0,-2) -- (4.1,2.1);
\draw [dashed, thick, blue] (0,-2) -- (-4.1,2.1);

\draw [<->,thick] (0,-3.5) -- (0,2.35);
\draw [<->,thick] (-4.5,-2) -- (4.5,-2);

\draw (-.35,2.5) node [scale = .85] {$t$};
\draw (4.75, -2.25) node [scale = .85] {$x^i$};
\draw (-.25,-2.25) node [scale = .85] {$\ms{O}$};

\draw (2,-4) node [scale = .85] {\small{Minkowski space = $(M_\ext,\, g_\ext)$}};

\draw [thick] (-.98,2.1) -- (-1.5,-.5);
\draw [thick, densely dashed] (-1.5, -.5) -- (-2.125, -3.625);

\draw [->] [thick] (-3.35,0) arc [start angle=120, end angle=80, radius=75pt];
\draw (-4.9, 0) node [scale = .85]{\small{An incomplete}};
\draw (-4.675, -.5) node [scale = .85]{\small{timelike geodesic}};
\draw (-5.325, -1) node [scale = .85]{\small{in $(M,g)$}};

\draw [->] [thick] (1.5,2.8) arc [start angle=140, end angle=180, radius=60pt];
\draw (2.0,3.25) node [scale = .85]{\small{Milne universe = $(M,g)$}};


%


\end{tikzpicture}
\]
\captionsetup{format=hang}
\caption{\small{The singularity theorem in \cite{BGV} applies to the Milne universe. However any incomplete geodesic extends to a complete geodesic within Minkowski space.  }}\label{singularity theorem counterexample figure}
\end{figure}

Assuming an inflationary era, $a''(\tau) > 0$, then Friedmann's equation implies $\rho + 3p < 0$. Hence inflationary models do not satisfy the strong energy condition. Therefore the singularity theorems above no longer apply. New singularity theorems were sought that did not require the strong energy condition. This was done by Borde and Vilenkin \cite{Borde, BV} and others \cite{GLcosmo, Lesourd_2019}. Borde and Vilenkin found that some models of inflationary theory also violate the weak energy condition \cite{BVweakviolation}. Then Guth, Borde, and Vilenkin produced a singularity theorem \cite{BGV}, which showed that, even if the weak energy condition is violated, then one has past incompleteness. However their theorem only applies to inflating regions of a spacetime. For example, their theorem applies to the Milne universe (because their theorem only requires an averaged Hubble expansion condition), but the Milne universe isometrically embeds into Minkowski space which is geodesically complete. See Figure \ref{singularity theorem counterexample figure}.

\medskip

Note that the geodesic displayed in Figure \ref{singularity theorem counterexample figure} is not a geodesic of a comoving observer. All the comoving observers emanate from the origin $\ms{O}$ (see Figure \ref{comoving figure in intro}). Section \ref{horizon problem section} gives a derivation of this result.

\section{Miscellaneous Proofs}

\medskip
\medskip

\subsection{Proof of Proposition \ref{timelike is open prop}}\label{I+ open appendix}

\medskip

\begin{Def}
\emph{
The \emph{Minkowski} metric in $\R^{n+1}$ is $\eta = \eta_{\mu\nu}dx^\mu dx^\nu = -(dx^0)^2 + \delta_{ij}dx^idx^j$. For $0 < \e < 1$, we define the \emph{narrow} and \emph{wide} Minkowski metrics 
\[
\eta^\e \,=\, -\frac{1 - \e}{1+\e}(dx^0)^2 + \delta_{ij}dx^idx^j \,\,=\,\, \eta + \frac{2\e}{1+\e}(dx^0)^2
\]
\[
\eta^{-\e}  \,=\, -\frac{1 + \e}{1-\e}(dx^0)^2 + \delta_{ij}dx^idx^j \,\,=\,\,  \eta - \frac{2\e}{1-\e}(dx^0)^2
\]
}
\end{Def}

\medskip
\medskip 

\begin{lem}\label{coord lem} Fix $k \geq 0$. Let $(M,g)$ be a $C^k$ spacetime. Fix $p \in M$. For any $0< \e < 1$ there is a coordinate system $\phi \colon U_\e \to \R^{n+1}$ with the following properties 
\begin{itemize}
\item[\emph{(1)}] $\phi(p) \, = \, 0$

\item[\emph{(2)}] $g_{\mu\nu}(p) \, =\, \eta_{\mu\nu}$


\item[\emph{(3)}] $I^+_{\eta^{\e}} (p, U_\e) \,\subset\, I^+ (p, U_\e) \,\subset\, I^+_{\eta^{-\e}} (p, U_\e)$.

\end{itemize}
Moreover if $\g \colon [a,b] \to M$ is a unit timelike curve with $\g(b) = p$, then we can choose the coordinate system so that $\phi \circ \g(t) = (t-b, 0, \dotsc, 0)$.

\end{lem}

\medskip

\proof

Pick a coordinate system $\phi \colon U \to \R^{n+1}$ with $\phi(p) = 0$ and apply Gram-Schmidt to obtain (2).
By continuity of the metric, given any $\e' > 0$, we can shrink our neighborhood so that $|g_{\mu\nu}(x)  - \eta_{\mu\nu}| < \e'$. Let $X = X^\mu \pd_{\mu}$ be a unit tangent vector (i.e. $|X^0| = 1$). Then
\begin{align*}
g(X,X) \,&<\,
\eta(X,X) + \e'\sum_{\mu,\, \nu =0}^n |X^\mu X^\nu| 
\\
&=\, \eta^\e(X,X) - \frac{2\e}{1+\e} + \e'\sum_{\mu,\, \nu = 0}^n|X^\mu X^\nu| \\
&=\, \eta^\e(X,X) - \frac{2\e}{1+\e} + \e'\left[1 + 2\sum_{i=1}^n |X^i| + \sum_{i,\,j = 1}^n|X^i X^j| \right] 
\end{align*}
If $X$ is $\eta^{\e}$-timelike, then $|X^i|^2/|X^0|^2 < (1-\e)/(1+\e)$. Since $|X^0| = 1$, we have
\begin{align*}
g(X,X) \,<\, \eta^\e(X,X) - \frac{2\e}{1+\e} + \e'\left[1 + 2n\sqrt{\frac{1-\e}{1+\e}} +n^2 \frac{1 - \e}{1+\e} \right].
\end{align*}
 By taking $\e' > 0$ small enough, we can ensure $2\e/(1+\e)$ is larger than the bracket term. This proves the first inclusion in (3). The proof of the second is analogous.

Now let $\g \colon [a,b] \to M$ be a unit timelike curve with $\g(b) = p$. Let $(y^0, y^1, \dotsc, y^n)$ be the coordinates on $U$ (i.e. $y^\mu = \pi^\mu \circ \phi$ where $\pi^\mu \colon \R^n \to \R$ are the canonical projections). Since $g_{\mu\nu}(p) = \eta_{\mu\nu}$, we can shrink $U$ so that $y^0$ is a time function (i.e. $\nabla y^0$ is past-directed timelike).  Since the definition of a timelike curve requires $\lim_{t \nearrow b}\g'(t)$ to be future-directed timelike, the function $(y^0 \circ \g)'(t)$ approaches a nonzero number as $t \nearrow b$. Therefore the inverse function theorem guarantees an interval $(b - \delta, \,b + \delta)$ around $b$ and a diffeomorphism $f\colon (b - \delta,\, b+\delta) \to (-\delta', \delta')$ such that $f = y^0 \circ \g$ on $(b-\delta,\,b]$. Let $U' \subset U$ be the preimage of $(-\delta', \delta)$ under $y^0$. We define new coordinates $(x^0, x^1, \dotsc, x^n)$ on $U'$ by 
\[
x^0(q) \,=\, y^0(q) \:\:\:\: \text{ and } \:\:\:\: x^i(q) = y^i(q) - y^i\big(\g \circ f^{-1}\circ y^0(q) \big).
\]
With these coordinates we have
\[
\pushQED{\qed}
x^0 \circ \g(t) \,=\, t-b \:\:\:\: \text{ and } \:\:\:\: x^i \circ \g(t) \,=\, 0. \qedhere
\popQED
\]

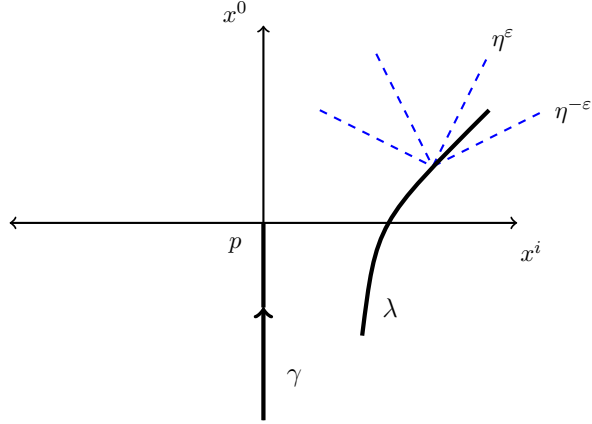
\begin{figure}[h]
\[
\begin{tikzpicture}[scale = 0.75]

\draw [dashed, thick, blue](1,2) -- (3,1) -- (5,2);
\draw [dashed, thick, blue](2,3) -- (3,1) -- (4,3);

\draw [->,thick] (0,-3.5) -- (0,3.5);
\draw [<->,thick] (-4.5,0) -- (4.5,0);
\draw [->,ultra thick](0, -3.5) -- (0, -1.5);
\draw[-, ultra thick](0, -1.5) -- (0,0);

\draw (-0.5,-0.4) node [scale = .85] {$p$};

\draw (-0.5,3.75) node [scale = .85] {$x^0$};
\draw (4.75,-0.5) node [scale = .85] {$x^i$};

\draw (4.25,3.25) node [scale = .85] {$\eta^\e$};
\draw (5.5,2) node [scale = .85] {$\eta^{-\e}$};

\draw [ultra thick, black] (1.75,-2) .. controls (2,0) .. (4,2);

\draw (2.25,-1.5) node [scale = .85] {\large{$\lambda$}};

\draw (0.55,-2.75) node [scale = .85] {\large{$\g$}};

\end{tikzpicture}
\]
\captionsetup{format=hang}
\caption{\small{The coordinate system appearing in Lemma \ref{coord lem}. The point $p$ is located at the origin where the metric is exactly Minkowski (i.e.  $g_{\mu\nu}(p) = \eta_{\mu\nu}$). The timelike curve $\g$ makes up the negative $x^0$-axis. Any timelike curve $\lambda \subset U_\e$ will always be $\eta^{-\e}$-timelike but it may be $\eta^\e$-spacelike.}}
\end{figure}

\medskip
\medskip

\noindent\underline{Proof of Proposition \ref{timelike is open prop}}:
\medskip

Fix $q \in I^+(p, U)$ and let $\g \colon [a,b] \to U$ be a timelike curve with $\g(a) = p$ and $\g(b) = q$. By rescaling we can assume $\g$ is a unit timelike curve. Let $\phi \colon U_\e \to \R^{n+1}$ be a coordinate system from Lemma \ref{coord lem} centered around $q$. Choose $\e = 3/5$ so that $\eta^\e$ has lightcones with slope 2. Choose $t < b$ such that $\g(t) \in U_\e$. Then $I^+_{\eta^\e}\big(\g(t), U_\e \cap U\big)$ is an open set since it's just the interior of a cone intersected with an open set. Moreover, it's contained in $I^+\big(\g(t), U_\e \cap U\big)$ which is contained in $I^+(p,U)$. 
\qed

\medskip
\medskip

\subsection{Proof of Lemma \ref{curvature sing lem}}\label{C2 appendix}

\medskip

Fix $k \geq 2$. Let $(M,g)$ be a $C^k$ Milne-like spacetime with a scale factor whose second derivative satisfies $a''(\tau) = \a \tau + C \tau^3 + o(\tau^3)$ where $\a, C \in \R$. Recall the coordinate system $\zeta = (t,r,\theta, \phi)$ from the proof of Theorem \ref{extension thm}. Here $t$ and $r$ are given by
\begin{equation}
t(\tau, R) \,=\, b(\tau)\cosh(R ) \:\:\:\: \text{ and } \:\:\:\: r(\tau, R) \,=\, b(\tau)\sinh(R)
\end{equation}
where $b \colon I \to (0,\infty)$ is given by $b(\tau) = \exp\left(\int_{\tau_0}^\tau \frac{1}{a(s)}ds \right)$ for some $\tau_0 > 0$. With respect to the coordinate system $\zeta = (t, r, \theta, \phi)$, the metric takes the form 
\begin{align}
g \,=\, \Omega^2\big(\tau(t,r)\big)\big[-dt^2 + dr^2 + r^2(d\theta^2 + \sin^2\theta d\phi^2)\big] 
\end{align}
where $\Omega = 1/b' = a/b$. We have to show for any $t_0 \geq 0$ the limits
\begin{align*}
\lim_{(t,r) \to (t_0, t_0)} \frac{\pd \Omega}{\pd t}(t,r) \:\:\:\: \:\:  \lim_{(t,r) \to (t_0, t_0)} \frac{\pd \Omega}{\pd r}(t,r) \:\:\:\:\:\: \lim_{(t,r) \to (t_0, t_0)} \frac{\pd^2 \Omega}{\pd t^2}(t,r) \:\:\:\: \:\: \lim_{(t,r) \to (t_0, t_0)} \frac{\pd^2 \Omega}{\pd r^2}(t,r)
\end{align*}
exist and are finite. Note that $t$ and $r$ appearing in the limits above are defined on the open set $U = \big\{(t, r, \theta, \phi) \mid t^2 - r^2 < b^2(\tau_{\rm max})  \text{ and } t > 0 \text{ and } r \geq 0\big\}$ and where $\tau_{\rm max} \in (0, +\infty]$ is given from the interval $I = (0, \tau_{\rm max})$ of the scale factor.

Note that $b$ is a strictly increasing $C^1$ function which is never zero. Therefore it is invertible and the derivative of its inverse is $(b^{-1})'\big(b(\tau)\big) = 1/b'(\tau)$. Recall $\tau = b^{-1}\big(\sqrt{t^2 - r^2} \big)$. Therefore $\pd \tau/ \pd t = t/(b'b)$. Since $\Omega = a/b = 1/b'$, the chain rule gives 
\begin{equation}
\frac{\pd \Omega}{\pd t} \,=\, \Omega' \frac{\pd \tau}{\pd t} \,=\, \left(\frac{a' - 1}{b} \right)\left( \frac{t}{b' b}\right) 
\end{equation}
Let's simplify notation by letting $a(\tau) = \tau + f(\tau)$. Then we have
\begin{equation}\label{Omega one derivative}
\frac{\pd \Omega}{\pd t} \,=\,  \left(\frac{f'}{b^2b'}\right)t
\end{equation}
Taking another derivative, we get
\begin{align}
\frac{\pd^2 \Omega}{\pd t^2} \,&=\,t \frac{\pd}{\pd t} \left(\frac{f'}{b^2b'}\right) \,+\, \left(\frac{f'}{b^2b'}\right) \,=\, t\left(\frac{f'}{b^2b'}\right)'\frac{\pd \tau}{\pd t} \,+\, \left(\frac{f'}{b^2b'}\right)
\\
\,&=\, t \left[\frac{f''(b^2b') - f'\big(2b(b')^2 + b^2b''\big)}{b^4(b')^2} \right]\left(\frac{t}{b'b} \right) \,+\, \left(\frac{f'}{b^2b'}\right)
\\
\,&=\, t^2\left[\frac{f''}{b^3(b')^2} \,-\, \frac{2f'}{b^4b'} \,-\, \frac{f'b''}{b^3(b')^3} \right] \,+\, \left(\frac{f'}{b^2b'}\right)
\end{align}
Plugging $b'' \,=\, (b/a)' \,=\, (b'a - a'b)/a^2 \,=\, -bf'/a^2$ into the above expression gives
\begin{equation}\label{Omega two derivatives}
\frac{\pd^2 \Omega}{\pd t^2} \,=\, t^2\left[\frac{f''}{b^3(b')^2} \,-\, \frac{2f'}{b^4b'} \,+\, \frac{(f')^2}{a^2 b^2(b')^3} \right] \,+\, \left(\frac{f'}{b^2b'}\right)
\end{equation}
From the proof of Theorem \ref{extension thm}, for $\e = 1$ there exists a $\delta > 0$ such that for all $0 < \tau < \delta$, we have
\begin{equation}\label{b eq}
\left(\frac{\tau}{\tau_0} \right)\left( \frac{1 - \tau^{\e_0}}{1 + \tau_0^{\e_0}}\right)^{-1/\e_0} \, < \, b(\tau) \, < \, \left(\frac{\tau}{\tau_0} \right)\left( \frac{1 + \tau^{\e_0}}{1 + \tau_0^{\e_0}}\right)^{-1/\e_0}
\end{equation}
where $\e_0$ is given by $a(\tau) = \tau + o(\tau^{1 + \e_0})$. 
Since $f''(\tau) = \a \tau + C\tau^3 + o(\tau^3)$, we have $f'(\tau) = \frac{1}{2}\a \tau^2 + \frac{1}{4}C\tau^4 + o(\tau^4)$.  Using (\ref{b eq}) along with $b' = b/a$, we see that for small $\tau$, we have $b = \tau/\tau_0 + o(\tau)$ and $b'(\tau) = 1/\tau_0 + o(\tau)$. Therefore the squeeze theorem gives
\begin{equation}\label{lim eq 1}
\lim_{\tau \to 0} \left( \frac{f''}{b^3(b')^2} - \frac{2f'}{b^4b'}\right) \,=\, \frac{1}{2}C\tau_0^5.
\end{equation}
Note that we needed the asymptotic condition  $a''(\tau) = \a \tau + C\tau^3 + o(\tau^3)$ to get the above equality. Similarly, we have both of the limits
\begin{equation}\label{lim eq 2}
\lim_{\tau \to 0}\, \frac{(f')^2}{a^2b^2(b')^3} \:\:\:\:\:\: \text{ and } \:\:\:\:\:\: \lim_{\tau \to 0} \,\frac{f'}{b^2b'}
\end{equation}
exist and are both finite. Plugging in (\ref{lim eq 1}) and (\ref{lim eq 2}) into (\ref{Omega two derivatives}) and (\ref{Omega one derivative}), we see that for any $t_0 \geq 0$, we have that the limits
\begin{equation}
\lim_{(t,r) \to (t_0, t_0)} \frac{\pd\Omega}{\pd t}\:\:\:\:\:\:\lim_{(t,r) \to (t_0, t_0)} \frac{\pd^2 \Omega}{\pd t^2}
\end{equation}
exist and are finite. Similarly, we obtain the same conclusion for $\frac{\pd \Omega}{\pd r}$ and $ \frac{\pd^2 \Omega}{\pd r^2}$. This completes the proof of Lemma \ref{curvature sing lem}.

\medskip
\medskip

\subsection{Proof of Proposition \ref{EC prop}}\label{energy equiv appendix}

\medskip

Fix $k \geq 2$. Let $(M,g)$ be a $C^k$ spacetime. Recall the Einstein tensor is $G_{\mu\nu} = R_{\mu\nu} - \frac{1}{2}Rg_{\mu\nu}$. We say $(M,g)$ satisfies the \emph{weak energy condition} if $G_{\mu\nu}X^\mu X^\nu \geq 0$ for all timelike $X$, and the \emph{strong energy condition} if $R_{\mu\nu}X^\mu X^\nu \geq 0$ for all timelike $X$.

Let $(M,g)$ be an FLRW spacetime. Following \cite{ON} and \cite{Wald}, we define the energy density $\rho$ and pressure function $p$ in terms of the Einstein tensor. If $u = \pd / \pd \tau$ and $e$ is any unit spacelike vector orthogonal to $u$, then

\begin{equation}\label{Einstein tensor eq 1}
\rho \,=\, \frac{1}{8\pi} G_{\mu\nu} u^\mu u^\nu
\:\:\:\:\:\: \text{ and } \:\:\:\:\:\:
p \,=\, \frac{1}{8\pi} G_{\mu\nu} e^\mu e^\nu.
\end{equation}

\medskip
\medskip

Also, as a consequence of isotropy, we have $R_{\mu\nu}u^\mu e^\nu = 0$ (see Corollary 12.10 of \cite{ON}). Therefore

\begin{equation}\label{Einstein tensor eq 2}
G_{\mu\nu} u^\mu e^\nu \,=\, 0
\end{equation}

The following proof shows 

\begin{itemize}

\item[(a)] The weak energy condition is equivalent to $\rho \geq 0$ and $\rho + p \geq 0$.

\item[(b)] The strong energy condition is equivalent to $\rho + p \geq 0$ and $\rho + 3p \geq 0$. 

\end{itemize}

\medskip
\medskip

\noindent\underline{\emph{Proof of Proposition \emph{\ref{EC prop}}}}:

\begin{enumerate}

	\item[(a)] Suppose $(M,g)$ satisfies the weak energy condition. Then $\rho = \frac{1}{8\pi} G_{\mu\nu}u^\mu u^\nu \geq 0$. Now let $e$ be a unit spacelike vector orthogonal to $u$. Fix $\e > 0$ and let $X$ be the timelike vector given by $X^\mu = (1 + \e)u^\mu + e^\mu$. Then by equations (\ref{Einstein tensor eq 1}) and (\ref{Einstein tensor eq 2}), we have
	\[
	0 \,\leq\, \frac{1}{8\pi}G_{\mu\nu}X^\mu X^\nu \,=\, (1 + \e)^2\rho + p.
	\]
	Since this is true for all $\e > 0$, we have $\rho + P \geq 0$. 
		Conversely, suppose $\rho \geq 0$ and $\rho + p \geq 0$. Let $X$ be any timelike vector. Decompose $X^\mu = a u^\mu + b e^\mu$ where $e$ is a unit spacelike vector orthogonal to $u$. Then by equations (\ref{Einstein tensor eq 1}) and (\ref{Einstein tensor eq 2}), we have $\frac{1}{8\pi}G_{\mu\nu} X^\mu X^\nu = a^2 \rho + b^2 p$. We have two cases: (1) $p \geq 0$ and (2) $p < 0$. In the first case we have $G_{\mu\nu}X^\mu X^\nu = a^2 \rho + b^2 p \geq 0$. The inequality follows from $\rho \geq 0$. Thus the weak energy condition holds. Now consider the case $p < 0$. Then, since $\rho + p \geq 0$, we have 
	\[
	\frac{1}{8\pi}G_{\mu\nu}X^\mu X^\nu \,=\, a^2 \rho + b^2 p  \,\geq\, (- a^2 + b^2)p \,=\, p g_{\mu\nu} X^\mu X^\nu \,\geq\, 0.
	\]
The last inequality follows because $p < 0$ and $X$ is timelike.

	\item[(b)] Let $G = g^{\mu\nu}G_{\mu\nu} = 8\pi(-\rho + 3p)$. Then contracting the Einstein equation with $g^{\mu\nu}$ shows $G = -R$. Rearranging the Einstein tensor gives 
	\[
	R_{\mu\nu} \,=\, G_{\mu\nu} + \frac{1}{2}Rg_{\mu\nu} \,=\, G_{\mu\nu} + 4\pi(\rho - 3p)g_{\mu\nu}
	\]
	Assume the strong energy condition holds. Then using the above equation, we have $0 \leq R_{\mu\nu} u^\mu u^\nu = 8\pi \rho - 4\pi(\rho - 3p) = 4\pi (\rho + 3p)$ which establishes $\rho + 3p \geq 0$. To establish $\rho + p \geq 0$, fix $\e > 0$ and consider the timelike vector $X = (1 + \e) u + e$. Then using equations (\ref{Einstein tensor eq 1}) and (\ref{Einstein tensor eq 2}), we have 
	\[
	0 \,\leq\, R_{\mu\nu}X^\mu X^\nu \,=\, \big[ G_{\mu\nu} + 4\pi(\rho - 3p)g_{\mu\nu}\big]X^\mu X^\nu \,=\, 4\pi\big[(1 + \e)^2\rho + \rho - p + 3p(1+\e)^2\big].
	\]
	Since $\e > 0$ was arbitrary, we have $\rho + p \geq 0$. Conversely, suppose $\rho + p \geq 0$ and $\rho + 3p \geq 0$.  Let $X$ be any timelike vector. Decompose $X^\mu = a u^\mu + b e^\mu$. Then we have $\frac{1}{8\pi}G_{\mu\nu}X^\mu X^\nu = a^2 \rho + b^2 p$. Using this in our expression for $R_{\mu\nu}$, we have
	\[
	R_{\mu\nu}X^\mu X^\nu \,=\, G_{\mu\nu}X^\mu X^\nu + 4\pi(\rho - 3p)g_{\mu\nu}X^\mu X^\nu \,=\, 4\pi\big[a^2(\rho + 3p) \,+\, b^2(\rho - p) \big].
	\]
	There are two cases to consider: (1) $\rho - p > 0$ and (2) $\rho - p \leq 0$. In case (1), we immediately have $R_{\mu\nu}X^\mu X^\nu \geq 0$. Now consider case (2). Since $X$ is timelike, we have $-a^2 + b^2 < 0$. Therefore it suffices to show $(\rho - p)/ (\rho + 3p) < 1$. Hence it suffices to show $p > 0$.  Indeed $\rho + p \geq 0$ and $\rho - p \leq 0$ together imply $p \geq 0$.  	\qed
	
	\end{enumerate}

\medskip
\medskip

\newpage


\end{document}